\documentclass[12pt]{article}

\usepackage[margin=1in]{geometry}
\usepackage{setspace}

\usepackage{amsmath,amssymb,amsfonts,amsthm,amscd,mathtools,mathrsfs,bbm,bm}
\usepackage{latexsym}

\usepackage{graphicx}
\usepackage{epstopdf}
\usepackage{rotating}
\usepackage{array}
\usepackage{booktabs}
\usepackage{multirow}
\usepackage{bigstrut}
\usepackage{subcaption}
\usepackage{pdflscape}
\usepackage{placeins}
\usepackage{siunitx}
\usepackage{colortbl}

\usepackage{enumerate}
\usepackage{enumitem}
\usepackage{comment}
\setlist[enumerate]{label=(\roman*), leftmargin=*}
\setlist[itemize]{leftmargin=*}

\usepackage{algorithm}
\usepackage{algpseudocode}

\usepackage{natbib}
\usepackage{url}
\usepackage{xcolor}
\usepackage[colorlinks=true,linkcolor=blue,citecolor=blue,urlcolor=blue,filecolor=blue]{hyperref}
\usepackage{xr-hyper}

\numberwithin{equation}{section}

\newcommand{\tensor}[1]{\mathcal{#1}}
\newcommand{\tensormult}[2]{\times_{#1} #2}
\newcommand{\vecop}{\mathrm{vec}}
\newcommand{\Tr}{\mathrm{tr}}

\newcommand{\Var}{\mathrm{Var}}

\newcommand{\E}{\mathbb{E}}
\newcommand{\R}{\mathbb{R}}
\newcommand{\kron}{\otimes}
\newcommand{\trans}{^\top}

\newcommand{\norm}[1]{\left\lVert #1\right\rVert}
\newcommand{\normF}[1]{\left\lVert #1\right\rVert_F}
\newcommand{\normS}[1]{\left\lVert #1\right\rVert_S}
\newcommand{\lambdamin}{\lambda_{\min}}

\newcommand{\diag}{\mathrm{diag}}

\newcommand{\offdiag}{\operatorname{off}}
\newcommand{\vecoff}{\operatorname{vec}_{\operatorname{off}}}

\newcommand{\blind}{0}
\newcommand{\usesuppxref}{0}
\providecommand{\disableexternalxrefs}{0}
\makeatletter
\newcommand{\externaldocumentnobib}[1]{%
  \let\XR@org@bibcite\bibcite
  \def\bibcite##1##2{}%
  \externaldocument{#1}%
  \let\bibcite\XR@org@bibcite
}
\makeatother
\ifnum\disableexternalxrefs=0
  \ifnum\usesuppxref=1
    \externaldocumentnobib{supp}
  \fi
\fi

\newtheorem{theorem}{Theorem}[section]

\newtheorem{proposition}[theorem]{Proposition}
\newtheorem{corollary}[theorem]{Corollary}
\newtheorem{assumption}[theorem]{Assumption}

\theoremstyle{plain}
\newtheorem{remark}[theorem]{Remark}
\theoremstyle{remark}

\begin{document}
\def\spacingset#1{\renewcommand{\baselinestretch}{#1}\small\normalsize}
\spacingset{1}

\if1\blind
{
    \title{\bf Tensor-BEKK: Conditional Covariance Modeling and Inference for Tensor-Valued Time Series}
    \author{}
    \date{}
    \maketitle
    \thispagestyle{empty}
} \fi

\if0\blind
{
    \title{\bf Tensor-BEKK: Conditional Covariance Modeling and Inference for Tensor-Valued Time Series}

    \author{
        \begin{tabular}{c}
            Huan Gong\textsuperscript{1,2} \quad
            Feiyu Jiang\textsuperscript{3}
 \\[0.6cm]
            \parbox{13cm}{\centering
                \footnotesize
                \textsuperscript{1}College of Systems Engineering, National University of Defense Technology, Changsha, Hunan, China\\[0.1cm]
                \textsuperscript{2}National Key Laboratory of Digital Intelligent Modeling and Simulation, Changsha, Hunan, China\\[0.1cm]
                \textsuperscript{3}Department of Statistics and Data Science, School of Management, Fudan University, Shanghai, China
            }
        \end{tabular}
    }
    \date{}
    \maketitle
} \fi

\vspace{-7mm}

\begin{abstract}

Modern economic and financial data are increasingly organized as multiway arrays, with observations indexed simultaneously by factors such as geographic regions, industrial sectors, asset categories or other economic indicators. Representing such data as tensor-valued time series allows their intrinsic multiway structure to be explicitly preserved and modeled. Although substantial effort has been devoted to modeling the conditional mean of tensor-valued time series, comparatively less attention has been paid to their conditional covariance dynamics. The latter remains particularly challenging, as unrestricted multivariate covariance models involve many parameters and substantial computational cost.   To address these challenges, we propose the Tensor-BEKK (T-BEKK) model, a tensor-structured
BEKK model that retains the positive definite covariance recursion for the
vectorized process while imposing Kronecker structure on the intercept and on
the ARCH and GARCH coefficient matrices. The model reduces the parameter
dimension and gives a mode-specific interpretation of the covariance intercept,
ARCH effects, and GARCH persistence. We establish stationarity, identification, and the asymptotic properties of the Gaussian quasi-maximum likelihood estimator. We further develop mode-specific restricted score tests tailored to the tensor structure, inference procedures for nonzero spillover intensities within each mode, and a portmanteau diagnostic test based on quadratic form residuals. For higher-dimensional settings,
we also introduce the Tensor-Factor-BEKK (TF-BEKK) model. Under a first-step negligibility
condition, its feasible second-step QMLE is asymptotically equivalent to the
oracle QMLE based on the latent factors. Simulations and two empirical
applications, covering currency futures and Chinese equity tensor portfolio
allocation, illustrate the finite-sample behavior and empirical usefulness of the
proposed methods.

\end{abstract}

\newpage
\spacingset{1.5} 

\section{Introduction}
In many economic and financial applications, time series observations are indexed by multiple cross-sectional characteristics and therefore have matrix- or tensor-valued structures rather than simple vector forms. Typical examples arise in financial markets, where the dynamics of assets such as stocks, bonds, and foreign exchange products are jointly influenced by multiple factors, including geographic regions, industrial sectors,  and asset categories. The interactions among these characteristics are often nonlinear and time-varying, making them difficult to characterize using conventional vector-valued models. Compared with traditional vector representations, matrix- and tensor-valued representations preserve the multiway structure of the data and allow interactions across different modes to be modeled explicitly, thereby giving clearer interpretation and greater modeling flexibility; see  e.g., \citet{wang2019matrixfactor}, \citet{chen2021mar}, \citet{chen2022tensor}, \citet{han2024simultaneous,han2024cp,chang2023matrixcp,chang2024identification}.

Traditional approaches for handling time series data with multiway structures can generally be divided into three categories. The first approach ignores such structure and simply  vectorizes the tensor process by writing $\bm x_t=\vecop(\tensor X_t)$, where  $\tensor X_t\in\R^{d_1\times\cdots\times d_m}$ is an order-$m$ tensor. The vectorized process $\bm x_t$ then has    dimension $d=\prod_{k=1}^m d_k$. In this form, any multivariate volatility specification for $\bm x_t$ must be able to deal with the ambient dimension $d$, and the difficulty becomes particularly severe when modeling the full conditional covariance matrix. From a statistical perspective, the estimation procedure must handle a large number of parameters, typically of order $O(d^2)$. This  necessitates additional structural assumptions such as sparsity \citep{basu2015regularized}, low-rank or factor structure \citep{bai2002determining,lam2012factor} or simplified parameterization \citep{engleledoitwolf2019}. From a computational perspective, likelihood-based methods repeatedly update and invert $d\times d$ covariance matrices, leading to substantial computational burdens in high-dimensional settings. Moreover, unrestricted vector models do not use the mode structure to parameterize and interpret volatility transmission. The second approach is to partition the data into several groups according to structural factors, such as geographic regions, industrial sectors, or asset categories, and then model each group separately or through blockwise dependence structures; see, e.g. \cite{englekelly2012,tong2024clustergarch,zhang2024factor}.  While such methods can partially alleviate the dimensionality issue and improve interpretability, they typically rely on prespecified grouping structures and may fail to capture interactions and volatility spillovers across different groups or tensor modes.
Another approach is to incorporate structural factors directly into the modeling framework, such as ARMA-X  \citep{hannan1976identification} or GARCH-X \citep{han2014asymptotic} models. In such approaches, structural factors  are typically encoded as covariates and introduced into the conditional mean or volatility models. While this strategy can partially account for heterogeneity across observations, it treats these structural labels as auxiliary explanatory variables and does not explicitly preserve the matrix or tensor structure of the data.
These limitations motivate matrix- and tensor-valued time series models that use the underlying multiway structure of the data to specify parsimonious covariance dynamics with a mode-specific interpretation.

Matrix- and tensor-valued time series analysis has become an active research area in statistics and econometrics in recent years; see \cite{bolivar2026analysis} for a recent review. Representative examples include matrix autoregressive models \citep{chen2021mar,jiang2025regularized}, two-way dynamic factor models \citep{yuan2023twoway}, matrix factor models \citep{wang2019matrixfactor,gao2025denoising}, structured low-rank matrix decompositions \citep{chang2023matrixcp}, tensor autoregressive models \citep{li2021tensorar,wang2024tensortar}, and tensor factor models \citep{chen2022tensor,han2024iterative,han2024cp, barigozzi2025tailrobust}. The aforementioned works primarily focus on conditional mean dynamics, low-rank dependence structures and latent factor modeling.  However, as pointed out by \citet{yu2025matrix}, the conditional mean signal in matrix- and tensor-valued financial time series is often relatively weak, while conditional heteroskedasticity and volatility clustering remain dominant stylized features  \citep{engle1982}. Conditional covariance and volatility dynamics are therefore central to the risk dependence structure of matrix- and tensor-valued processes.

Since the pioneering works of \citet{engle1982} and \citet{bollerslev1986}, the generalized autoregressive conditional heteroskedasticity (GARCH)  models   have become a fundamental framework for modeling time-varying volatility and risk dynamics in economics and finance. Extending these ideas to multivariate settings, \citet{engle1995} introduced the BEKK parameterization, which ensures the positive definiteness of the conditional covariance matrix through a recursive matrix formulation. Comprehensive reviews of multivariate GARCH and BEKK-type models can be found in \citet{bauwens2006}, \citet{silvennoinen2009}, and \citet{francq2019garch}, while their stationarity and Gaussian quasi-maximum likelihood estimation (QMLE) have been studied by \citet{jeantheau1998}, \citet{comtelieberman2003}, and \citet{boussama2011}, among others.

To address the curse of dimensionality in multivariate volatility modeling, existing studies have proposed various structural simplifications, including dynamic correlation models \citep{tsetsui2002,engle2002dcc,aielli2013}, factor-based covariance and volatility structures \citep{lanne2007,engleledoitwolf2019}, and correlation constraints such as dynamic equicorrelation \citep{englekelly2012}. Among these approaches, the BEKK framework is useful because it directly specifies a positive definite recursion for the full conditional covariance matrix. 

Nevertheless, the existing multivariate volatility literature has been developed almost exclusively for vector-valued processes and provides only limited tools for matrix- or tensor-valued time series. One notable exception is the Matrix-GARCH model of \citet{yu2025matrix}, which models volatility dynamics through a recursion at the matrix level and consequently induces a covariance updating mechanism for $\vecop(\tensor X_t)$ that differs from the BEKK-type recursion considered in this paper. In particular, \citet{yu2025matrix} proposed rowwise and columnwise BEKK-type specifications for matrix-valued volatility modeling, together with a trace-GARCH formulation to address identification issues. In contrast, the present paper develops a tensor-structured framework by imposing Kronecker structures directly on the BEKK coefficient matrices. Extending Matrix-GARCH-type constructions from matrices to general higher-order tensors is non-trivial because tensor-valued processes have several interacting modes.
Another related contribution is the tensor dynamic conditional correlation (TDCC) model proposed by \citet{yu2025tdcc}, which combines entrywise conditional variances with mode-specific conditional correlation matrices following DCC-type \citep{engle2002dcc} dynamics after standardization.  While this framework allows heterogeneous marginal volatility behavior and mode-specific correlation structures, \citet{yu2025tdcc} focus mainly on model construction, estimation, and numerical performance, with less emphasis on probabilistic properties of the model and related statistical inference procedures.

This paper proposes a tensor-structured BEKK model for the conditional covariance of tensor-valued time series. The key idea is to retain the classical BEKK recursion for the vectorized process while imposing a parsimonious Kronecker structure on the intercept matrix as well as the ARCH and GARCH coefficient matrices, thereby preserving the tensor structure and giving a direct mode-specific interpretation.
Our contributions are threefold. 

The first and main contribution is the Tensor-BEKK (T-BEKK) model. Starting from the classical BEKK$(1,1)$ recursion for $\bm x_t=\vecop(\tensor X_t)$, we impose a Kronecker product structure on the intercept and on the ARCH and GARCH coefficient matrices. This preserves the BEKK covariance recursion and its positive definite structure while reducing the parameter count from order $O(d^2)$ to order $O(\sum_{k=1}^m d_k^2)$ compared with the vector counterpart. The parameterization also gives a clear mode-specific interpretation: mode-specific factors govern the covariance intercept, ARCH effects, and GARCH persistence along each tensor mode, while their Kronecker products determine the overall covariance recursion.

Second, we develop the statistical theory and inference procedures for the T-BEKK model. We establish a sufficient condition for strict stationarity and ergodicity. Under an injectivity condition on the global BEKK representation, we identify the normalized Kronecker parameters. We further establish strong consistency and asymptotic normality of the Gaussian quasi-maximum likelihood estimator under standard BEKK-type regularity conditions. Building on this asymptotic theory, we construct mode-specific restricted score tests and spillover intensity measures for ARCH and GARCH volatility transmission, together with a consistent sandwich covariance estimator and an implementable portmanteau diagnostic test that accounts for parameter estimation.

Third, we develop a factor extension for higher-dimensional settings. Even under the
Kronecker restriction, evaluating the T-BEKK likelihood still requires repeated inversion of $d\times d$ covariance matrices, and can remain computationally demanding. We therefore introduce the Tensor-Factor-BEKK (TF-BEKK) model, in which a low-dimensional latent tensor follows its own T-BEKK recursion and the observed tensor is linked to the latent process through mode-specific loading matrices. We propose a two-step estimation procedure: in the first step, the loading spaces and latent tensor factors are estimated; in the second step, a T-BEKK model is fitted to the estimated latent factors by QMLE.  Under the stated oracle regularity conditions, the feasible second-step QMLE is asymptotically equivalent to the oracle QMLE based on the oriented latent factors when the first-step estimation is sufficiently accurate.  Finally, simulation evidence and two empirical applications, covering currency futures and Chinese equity tensor portfolio allocation, illustrate the finite-sample behavior and empirical usefulness of the proposed framework.

The remainder of the paper is organized as follows. Section~\ref{sec:model} introduces the T-BEKK model and studies its structural and identification properties. Section~\ref{sec:qmle} develops the Gaussian QMLE and its large-sample theory. Section~\ref{sec:spillover} studies mode-specific volatility spillover inference, and Section~\ref{sec:checking} presents a portmanteau diagnostic test based on quadratic form residuals. Section~\ref{sec:factor} introduces the TF-BEKK extension and studies the associated two-step estimator. Section~\ref{sec:simulation} reports Monte Carlo evidence on the finite-sample behavior of the QMLE, the mode-specific spillover tests, and the portmanteau test. Section~\ref{sec:application} presents the two empirical applications. Section~\ref{sec:conclusion} concludes. All proofs are collected in the supplement.

\medskip
\noindent\textit{Notation.}
Throughout, $\R=(-\infty,\infty)$ and $\R_+=[0,\infty)$. Calligraphic symbols such as $\tensor X_t$ denote tensors, bold lowercase symbols such as $\bm x_t$ denote vectors, and uppercase symbols such as $H_t$, $A$, and $B$ denote matrices. We write $I_p$ for the $p\times p$ identity matrix; $\xrightarrow{p}$ and $\xrightarrow{d}$ denote convergence in probability and in distribution, respectively. For a square matrix $X$, $\Tr(X)$, $\det(X)$, and $\rho(X)$ denote its trace, determinant, and spectral radius. In addition, we write
$\offdiag(X)=X-\diag(X)$, where $\diag(X)$ denotes the diagonal matrix formed
from the diagonal entries of $X$. For a rectangular matrix $X$, $X\trans$, $\vecop(X)$, $\normF{X}$, and $\normS{X}$ denote its transpose, columnwise vectorization, Frobenius norm, and spectral norm. For an order-$m$ tensor $\tensor X\in\R^{d_1\times\cdots\times d_m}$, $\vecop(\tensor X)$ denotes the vector in $\R^{\prod_{k=1}^m d_k}$ obtained by stacking the mode-1 fibers in the order of modes $2,\dots,m$. The Kronecker product is written as $\kron$, and $\mathcal{O}(k):=\{R\in\R^{k\times k}:R\trans R=RR\trans=I_k\}$ denotes the group of $k\times k$ orthogonal matrices.

\section{The T-BEKK model}
\label{sec:model}

\subsection{Model specification}
Let $\{\tensor X_t\}_{t\in\mathbb Z}$ be a centered order-$m$
tensor-valued time series with $\tensor X_t\in\R^{d_1\times\cdots\times d_m}$
satisfying $\E(\tensor X_t\mid \mathcal F_{t-1})=0$, where $\mathcal{F}_t=\sigma(\tensor X_t, \tensor X_{t-1},\ldots)$ is the natural filtration. Write
$\bm x_t=\vecop(\tensor X_t)\in\R^d$, where $d=\prod_{k=1}^m d_k$, and let
\begin{equation*}
H_t=\Var(\bm x_t\mid \mathcal F_{t-1}),
\end{equation*}
denote its conditional covariance matrix. Our goal is to model $H_t$ in a
way that preserves the positive definite BEKK recursion while using the
multidimensional array structure of the data. Following \cite{engle1995}, we write
\begin{flalign}\label{eq:tbekk-vec}
\begin{split}
\bm x_t=&H_t^{1/2}\bm\eta_t,\\
H_t=&\Omega + A\bm x_{t-1}\bm x_{t-1}\trans A\trans + B H_{t-1} B\trans,
\end{split}
\end{flalign}
where $\{\bm\eta_t\}$ is an i.i.d.\ innovation sequence such that
$\bm\eta_t$ is independent of $\mathcal{F}_{t-1}$ for each $t$ and satisfies
$\E(\bm\eta_t)=\bm 0$ and $\E(\bm\eta_t\bm\eta_t\trans)=I_d$, and the three $d\times d$ coefficient matrices admit the Kronecker product
representations
\begin{equation}
\Omega = \Omega_m\kron\cdots\kron\Omega_1,
\qquad
A=A_m\kron\cdots\kron A_1,
\qquad
B=B_m\kron\cdots\kron B_1.
\label{eq:kron-params}
\end{equation}
Here, $\Omega_k\in\R^{d_k\times d_k}$ is a positive definite  intercept matrix, and 
$A_k,B_k\in\R^{d_k\times d_k}$ capture the ARCH and GARCH effects for mode $k$, respectively. We refer to $\Omega$, $A$, and $B$ as the
global BEKK coefficient matrices, and to
$\{\Omega_k,A_k,B_k\}_{k=1}^m$ as their mode-specific counterparts. We call \eqref{eq:tbekk-vec} together with \eqref{eq:kron-params} the T-BEKK model. 

When
$m=1$, the Kronecker representation is trivial and T-BEKK reduces to
the classical BEKK$(1,1)$ model. More generally, for a fixed mode partition
$(d_1,\dots,d_m)$, T-BEKK is the subclass of vector BEKK$(1,1)$ models whose
intercept and ARCH/GARCH coefficient matrices each admit the single
Kronecker product representation in \eqref{eq:kron-params}. 

A useful feature of the Kronecker specification in \eqref{eq:kron-params} is that the ARCH component can be represented through multilinear
transformations of the tensor process. In particular, the standard mode-$k$ product $\times_k$ is defined as follows: for $\tensor X\in\R^{d_1\times\cdots\times d_m}$ and $M\in\R^{\widetilde d_k\times d_k}$, the tensor $\tensor X\times_k M$ lies in
$\R^{d_1\times\cdots\times d_{k-1}\times \widetilde d_k\times d_{k+1}\times\cdots\times d_m}$ and has entries \citep{kolda2009tensor}
\[
(\tensor X\times_k M)_{i_1\cdots i_{k-1} j i_{k+1}\cdots i_m}
=
\sum_{i_k=1}^{d_k} x_{i_1\cdots i_k\cdots i_m} m_{j i_k}.
\]
If we define
\begin{equation}
\tensor Y_{t-1}=\tensor X_{t-1}\tensormult{1}{A_1}\cdots\tensormult{m}{A_m},
\label{eq:Ytdef}
\end{equation}
then $\vecop(\tensor Y_{t-1})=A\bm x_{t-1}$ and \eqref{eq:tbekk-vec} becomes
\begin{equation}
H_t=\Omega+\vecop(\tensor Y_{t-1})\vecop(\tensor Y_{t-1})\trans + B H_{t-1} B\trans.
\label{eq:tbekk-tensorform}
\end{equation}

\begin{remark}
Under the vectorization order fixed in \eqref{eq:kron-params}, let
$\nu(i_1,\dots,i_m)$ denote the coordinate associated with tensor cell
$(i_1,\dots,i_m)$:
\[
\nu(i_1,\dots,i_m)
=1+\sum_{k=1}^m (i_k-1)\prod_{\ell=1}^{k-1} d_\ell,
\]
with the convention $\prod_{\ell=1}^0=1$. Then
$A_{\nu(i_1,\dots,i_m),\,\nu(j_1,\dots,j_m)}
=\prod_{k=1}^m (A_k)_{i_kj_k}$, with analogous relations for $B$ and
$\Omega$. Thus links between tensor cells decompose into mode-specific
contributions: $A_k$ governs shock propagation along mode $k$, whereas $B_k$
governs the propagation of past conditional covariance along that mode.
\end{remark}

The full matrix $H_t$ also induces mode-specific marginal conditional
covariances. Let $X_{t,(k)}\in\R^{d_k\times d_{-k}}$ be the mode-$k$ unfolding
of $\tensor X_t$, where $d_{-k}=\prod_{\ell\ne k}d_\ell$, and define
\begin{equation}
\Sigma_{k,t}
=
\E\{X_{t,(k)}X_{t,(k)}\trans\mid\mathcal F_{t-1}\}.
\label{eq:modewise-cov-Sigmak}
\end{equation}
Equivalently, $\Sigma_{k,t}$ is the partial trace of $H_t$ over all modes
except mode $k$:
\[
(\Sigma_{k,t})_{ij}
=
\sum_{a_{-k}}
\{H_t\}_{\nu(i,a_{-k}),\,\nu(j,a_{-k})},
\]
where $a_{-k}=(a_1,\dots,a_{k-1},a_{k+1},\dots,a_m)$ and
$\nu(i,a_{-k})$ denotes the vectorization index obtained by placing $i$ in
the $k$th coordinate. For the formula below, let $A_{-k}$ denote the Kronecker
product of $\{A_\ell:\ell\ne k\}$ in the column order of the mode-$k$
unfolding, so that
\[
Y_{t-1,(k)}=A_kX_{t-1,(k)}A_{-k}\trans,
\]
where $Y_{t-1,(k)}$ denotes the mode-$k$ unfolding of
$\tensor Y_{t-1}$ defined in \eqref{eq:Ytdef}. The GARCH term involves the
weighted partial trace $\mathscr W_{k,B}(H_{t-1})$, defined entrywise by
\[
\{\mathscr W_{k,B}(H_{t-1})\}_{ij}
=
\sum_{a_{-k},b_{-k}}
\left\{\prod_{\ell\ne k}
(B_\ell\trans B_\ell)_{a_\ell b_\ell}\right\}
 \{H_{t-1}\}_{\nu(i,a_{-k}),\,\nu(j,b_{-k})}.
\]
 
\begin{proposition}
\label{prop:modewise-cov}
Under the T-BEKK recursion \eqref{eq:tbekk-vec}--\eqref{eq:kron-params}, the
mode-$k$ conditional covariance matrix in \eqref{eq:modewise-cov-Sigmak}
satisfies
\begin{equation}
\Sigma_{k,t}
=
c_{\Omega,k}\Omega_k
+
A_kX_{t-1,(k)}
(A_{-k}\trans A_{-k})
X_{t-1,(k)}\trans A_k\trans
+
B_k\,\mathscr W_{k,B}(H_{t-1})\,B_k\trans,
\label{eq:modewise-cov-formula}
\end{equation}
where
$
c_{\Omega,k}=\prod_{\ell\ne k}\Tr(\Omega_\ell).
$
\end{proposition}

Proposition \ref{prop:modewise-cov} decomposes the mode-$k$ conditional covariance into three components corresponding to the intercept, ARCH effect, and GARCH persistence, respectively. The last term generally depends on the full
matrix $H_{t-1}$ through $\mathscr W_{k,B}(H_{t-1})$; it can be reduced to
$B_k\Sigma_{k,t-1}B_k\trans$  under additional restrictions such as
$B_\ell\trans B_\ell=I_{d_\ell}$ for all $\ell\ne k$. 

Relative to an unrestricted vector BEKK specification,
\eqref{eq:kron-params} reduces the number of free parameters from order
$O(d^2)$ to order $O(\sum_{k=1}^m d_k^2)$, yielding a more parsimonious covariance recursion.  For example,  consider the matrix setting with $m=2$, where the return matrix is indexed by geographic region and asset category.   The matrix $A_1$ governs ARCH effects along the geographic mode, while $A_2$ governs their propagation along the asset category mode. The effect of past return shocks on the conditional covariance dynamics can be interpreted as a separable interaction between regional and asset category effects. This mode-specific interpretation helps distinguish different structural sources of risk.

\subsection{Stationarity}
In this subsection, we study the structural properties of the T-BEKK model. Note that the BEKK form of \eqref{eq:tbekk-vec} preserves positive definiteness:
if $\Omega$ is positive definite, then $H_t$ is positive definite for all
$t$, because the ARCH and GARCH terms are positive semidefinite. 
The next proposition gives the resulting stationarity and ergodicity of the model.

\begin{proposition}
\label{prop:stationarity}
If the distribution of $\bm\eta_t$ admits a density that is positive on a neighborhood of $\bm 0$ with respect to the
Lebesgue measure, and
\begin{equation}
\rho\!\left(A\kron A + B\kron B\right)<1,
\label{eq:stationarity}
\end{equation}
where $\rho(\cdot)$ denotes the spectral radius, then the T-BEKK recursion
\eqref{eq:tbekk-vec} admits a unique strictly stationary and ergodic solution.
Moreover, this solution satisfies
$\E\|\bm x_t\|^2<\infty$.
\end{proposition}

The Kronecker restriction narrows the admissible parameter space but does not alter the
recursion itself, so the sufficient condition for the classical BEKK model can still be applied. Proposition 2.3 is hence an immediate corollary of 
\citet[Theorem~2.4]{boussama2011}, see also \citet[Theorem~10.5]{francq2019garch}.  
Under \eqref{eq:stationarity}, iterating the recursion
yields the following representation for the unique
strictly stationary solution:
\begin{equation}
H_t = \sum_{j=0}^{\infty} B^j\Omega (B\trans)^j + \sum_{j=1}^{\infty} B^{j-1}A\bm x_{t-j}\bm x_{t-j}\trans A\trans (B\trans)^{j-1}.
\label{eq:infinite-rep}
\end{equation}

\subsection{Identification}
The Kronecker specification in \eqref{eq:kron-params} is not unique. For example, the model remains unchanged if one replaces $\Omega_i$ and $\Omega_j$ by $\widetilde{\Omega}_i=c\Omega_i$ and $\widetilde{\Omega}_j=c^{-1}\Omega_j$, respectively, for any $c>0$ and $i\neq j$. The same scaling indeterminacy also applies to the Kronecker factors associated with $A$ and $B$. Identification in the present setting therefore involves two distinct aspects: normalization of the Kronecker representation itself, and injectivity of the mapping from the tensor-structured parameters to the induced global BEKK recursion. The following assumption separates these two roles.

\begin{assumption}
\label{ass:id}
The following conditions hold.
\begin{enumerate}
    \item For each $k=1,\dots,m$, $\Omega_k=C_kC_k\trans$ with $C_k$ lower triangular and $(C_k)_{ii}>0$ for all $i$.
    \item $(C_k)_{11}=1$, $(A_k)_{11}=1$, and $(B_k)_{11}=1$ for $k=2,\dots,m$, and $(A_1)_{11}>0$, $(B_1)_{11}>0$.
    \item If $H_t(\bm\theta)=H_t(\bm\theta_0)$ almost surely for all $t$, then $\Omega(\bm\theta)=\Omega(\bm\theta_0)$, $A(\bm\theta)=A(\bm\theta_0)$, and $B(\bm\theta)=B(\bm\theta_0)$.
\end{enumerate}
\end{assumption}

Assumption \ref{ass:id}
\upshape(i)--\upshape(ii) fix the Cholesky representation and remove the scale
and sign indeterminacies of the Kronecker factors. Assumption \ref{ass:id}~\upshape(iii)
imposes injectivity at the level of the global BEKK representation, in the
spirit of standard identification conditions for BEKK models
\citep{pedersen2014multivariate}.

\begin{remark}
\label{rem:id-sufficient}
Assumption \ref{ass:id}~  \upshape(iii) follows from standard identification conditions for the
corresponding vector BEKK representation; see, e.g.
\citet{engle1995}. Specifically, suppose that $\bm\eta_t$ admits a density
that is positive on a nonempty open subset of $\mathbb R^d$ and that
$A(\bm\theta_0)$ is nonsingular. The sign indeterminacies of $A$ and $B$ are
resolved by an appropriate normalization, such as requiring their $(1,1)$
entries to be positive. Under these conditions,
Assumption \ref{ass:id}~  \upshape(iii) holds directly.
\end{remark}

\begin{proposition}
\label{prop:id}
Under Assumption~\ref{ass:id}, if $H_t(\bm\theta)=H_t(\bm\theta_0)$ almost surely for all $t$, then $\bm\theta=\bm\theta_0$.
\end{proposition}

Proposition~\ref{prop:id} shows that, once the global BEKK matrices are identified from the conditional covariance process, the imposed normalizations uniquely determine their mode-specific Kronecker factors. This two-stage identification argument provides the foundation for the estimation and inference theory developed in Section~\ref{sec:qmle}.

\section{QMLE and asymptotic theory}
\label{sec:qmle}
This section develops the estimation and inference theory for the proposed model by deriving  the Gaussian QMLE for
$\bm\theta_0$ and establishing its consistency and asymptotic normality. Let $\bm\theta$ denote the vector
of free entries in $\{C_k,A_k,B_k\}_{k=1}^m$, let $\Theta$ denote the constrained parameter space defined by Assumption~\ref{ass:id} (i)--(ii), and
let $\{H_t(\bm\theta)\}$ denote the conditional covariance sequence generated by
\eqref{eq:tbekk-vec}.  
Note that throughout Sections~\ref{sec:qmle}--\ref{sec:checking}, the mode
dimensions $d_1,\dots,d_m$ are treated as fixed as $T\to\infty$. Accordingly,
$d=\prod_{k=1}^m d_k$ and the parameter dimension
$p=\dim(\bm\theta)$ are fixed in the asymptotic analysis below. 

\subsection{Gaussian quasi-maximum likelihood estimation}
For a given parameter vector $\bm\theta$, define the infeasible negative
Gaussian quasi-log-likelihood criterion, up to additive and positive
multiplicative constants, by
\begin{equation*}
L_T(\bm\theta)
=
\frac{1}{T}\sum_{t=1}^T \ell_t(\bm\theta),
\quad
\ell_t(\bm\theta)
=
\log\det\{H_t(\bm\theta)\}
+
\bm x_t\trans H_t(\bm\theta)^{-1}\bm x_t.
\end{equation*}
Because $H_t(\bm\theta)$ depends on the infinite past, we evaluate the
criterion using a recursion initialized at $\bm x_0=\bm 0$ and
$\widetilde H_0=I_d$. For any $\bm\theta\in\Theta$, define recursively \[ \widetilde H_t(\bm\theta) = \Omega(\bm\theta) + A(\bm\theta)\bm x_{t-1}\bm x_{t-1}\trans A(\bm\theta)\trans + B(\bm\theta)\widetilde H_{t-1}(\bm\theta)B(\bm\theta)\trans, \qquad t=1,\dots,T. \]
The resulting feasible negative Gaussian quasi-log-likelihood function is \[ \widetilde L_T(\bm\theta) = \frac{1}{T}\sum_{t=1}^T\widetilde\ell_t(\bm\theta), \qquad \widetilde\ell_t(\bm\theta) = \log\det\{\widetilde H_t(\bm\theta)\} + \bm x_t\trans\widetilde H_t(\bm\theta)^{-1}\bm x_t, \]
and the Gaussian QMLE is defined by \begin{equation} \widehat{\bm\theta}_T = \arg\min_{\bm\theta\in\Theta} \widetilde L_T(\bm\theta). \label{eq:qmle} \end{equation}

\subsection{Regularity conditions}
We next introduce regularity conditions used to establish
consistency and asymptotic normality. They are tensor analogues of the
standard QMLE conditions for multivariate GARCH and BEKK models
\citep{bollerslevwooldridge1992,jeantheau1998,comtelieberman2003,%
francq2019garch}, adapted to the Kronecker parameterization and the
identification restrictions in Assumption~\ref{ass:id}.

\begin{assumption}
\label{ass:innov}
The process
$\{\bm x_t\}$ is strictly stationary and ergodic with
$\E\norm{\bm x_t}^{s_x}<\infty$ for some $s_x>0$.
\end{assumption}

\begin{assumption}
\label{ass:ps}
The parameter space $\Theta$ is compact, and there exist $\delta>0$ and
$\gamma\in(0,1)$ such that
\[
\inf_{\bm\theta\in\Theta}\lambdamin\{\Omega(\bm\theta)\}\ge \delta,
\qquad
\sup_{\bm\theta\in\Theta}\rho\{B(\bm\theta)\kron B(\bm\theta)\}\le \gamma.
\]
\end{assumption}

\begin{assumption}
\label{ass:deriv}
There exists a neighborhood $N(\bm\theta_0)$ of $\bm\theta_0$ such that, for
each $t$, the map $\bm\theta\mapsto H_t(\bm\theta)$ is twice continuously
differentiable on $N(\bm\theta_0)$ almost surely. In addition, there exist exponents
$s>1$, $q>2$, and $r>2$ satisfying $2q^{-1}+2r^{-1}=1$ and
$s^{-1}+2r^{-1}=1$ such that, for all parameter indices $i,j$,
\begin{align*}
\E\sup_{\bm\theta\in N(\bm\theta_0)}\Big\|H_t(\bm\theta)^{-1/2}\frac{\partial^2H_t(\bm\theta)}{\partial\theta_i\partial\theta_j}H_t(\bm\theta)^{-1/2}\Big\|^s &<\infty,\\
\E\sup_{\bm\theta\in N(\bm\theta_0)}\Big\|H_t(\bm\theta)^{-1/2}\frac{\partial H_t(\bm\theta)}{\partial\theta_i}H_t(\bm\theta)^{-1/2}\Big\|^q &<\infty,\\
\E\sup_{\bm\theta\in N(\bm\theta_0)}\Big\|H_t(\bm\theta)^{-1/2}H_t(\bm\theta_0)^{1/2}\Big\|^r &<\infty.
\end{align*}
\end{assumption}
Assumption \ref{ass:innov} is mild and standard. Note that under the sufficient
stationarity condition in Proposition~\ref{prop:stationarity},
Assumption~\ref{ass:innov} holds automatically with $s_x=2$. Assumption~\ref{ass:ps} regulates the parameter space.
The lower eigenvalue bound prevents the conditional covariance matrices from
becoming nearly singular, whereas the spectral radius restriction ensures
uniform geometric stability of the covariance recursion and, consequently,
uniform control of the initialization error. 
Assumption~\ref{ass:deriv} controls both the local smoothness and the
integrability of the first- and second-order parameter derivatives of the
conditional covariance process, which are used to prove the
asymptotic normality of the QMLE. The following proposition provides primitive
sufficient conditions for Assumption~\ref{ass:deriv}, paralleling the moment
conditions commonly imposed in the classical BEKK literature; see, e.g. \citet{pedersen2014multivariate}.

\begin{proposition}
\label{prop:deriv_primitive}
Suppose Assumptions~\ref{ass:innov} and~\ref{ass:ps} hold and that $\bm\theta_0$ admits a
neighborhood $N_0$ whose closure is a compact subset of $\Theta$. If 
$\E\norm{\bm x_t}^6<\infty$, then Assumption~\ref{ass:deriv} holds with
$
(s,q,r)=\left(3/2,3,6\right).
$
\end{proposition}

\subsection{Consistency and asymptotic normality}
We next establish strong consistency of the QMLE and then derive
its   asymptotic normality.

\begin{theorem}
\label{th:consistency}
Under Assumptions~\ref{ass:id},~\ref{ass:innov}, and~\ref{ass:ps}, the QMLE in \eqref{eq:qmle} is strongly consistent: $\widehat{\bm\theta}_T\to\bm\theta_0$ almost surely as $T\to\infty$.
\end{theorem}
Theorem~\ref{th:consistency} extends the strong consistency result for the
classical BEKK QMLE \citep[][Section~10.4]{francq2019garch} to the normalized Kronecker parameterization. In
particular, Proposition~\ref{prop:id} ensures that identification of the global
BEKK matrices also identifies their mode-specific parameters.

Let
\[
\mathcal J_0=
\E\!\left\{
\frac{\partial^2\ell_t(\bm\theta_0)}
{\partial\bm\theta\partial\bm\theta^\top}
\right\},
\qquad
\mathcal I_0=
\E\!\left\{
\frac{\partial\ell_t(\bm\theta_0)}{\partial\bm\theta}
\frac{\partial\ell_t(\bm\theta_0)}{\partial\bm\theta^\top}
\right\}.
\]

\begin{theorem}
\label{th:an}
Suppose Assumptions~\ref{ass:id}-
\ref{ass:deriv} hold, $\bm\theta_0$ is an interior point of $\Theta$, and
$\E\|\bm\eta_t\|^4<\infty$. If $\mathcal J_0$ and $\mathcal I_0$ are positive
definite, then
\[
\sqrt T\,(\widehat{\bm\theta}_T-\bm\theta_0)
\xrightarrow{d}
\mathcal N\!\left(
\bm 0,\mathcal J_0^{-1}\mathcal I_0\mathcal J_0^{-1}
\right).
\]
\end{theorem}
The asymptotic covariance in Theorem~\ref{th:an} has the classical sandwich form
for Gaussian QMLE.   To make its components explicit, let  $
\mathcal D_t={\partial\vecop\{H_t(\bm\theta_0)\}}/{\partial\bm\theta\trans},$ $\Lambda_t=
\mathcal D_t^\top
\{H_t(\bm\theta_0)^{-1/2}\kron
H_t(\bm\theta_0)^{-1/2}\},$ and $\mathcal K=\E\{\vecop(I_d-\bm\eta_t\bm\eta_t\trans)\vecop(I_d-\bm\eta_t\bm\eta_t\trans)\trans\}$. Then 
\[
\mathcal J_0=
\E\!\left[
\mathcal D_t^\top
\{H_t(\bm\theta_0)^{-1}\kron H_t(\bm\theta_0)^{-1}\}
\mathcal D_t
\right],
\qquad
\mathcal I_0=
\E(\Lambda_t\mathcal K\Lambda_t^\top).
\]
Therefore, the effect of non-Gaussian innovations on the asymptotic covariance
is captured by the fourth-moment matrix $\mathcal K$. If $\bm\eta_t$ is Gaussian, then
$\mathcal I_0=2\mathcal J_0$ under the likelihood normalization adopted in
Section~\ref{sec:qmle}, and the limiting covariance reduces to
$2\mathcal J_0^{-1}$.

The proof of Theorem \ref{th:an} suggests that the QMLE  admits the Bahadur representation
\begin{equation}
\widehat{\bm\theta}_T-\bm\theta_0
=
\frac{1}{T}\mathcal J_0^{-1}
\sum_{t=1}^T
\Lambda_t\vecop(\bm\eta_t\bm\eta_t\trans-I_d)
+o_p(T^{-1/2}),
\label{eq:bahadur}
\end{equation}
which is used in Section~\ref{sec:checking} to account for
parameter estimation in the portmanteau statistic.

For feasible inference, define the Jacobian based on the initialized covariance
recursion by
$
\widetilde{\mathcal D}_t(\bm\theta)
=
{\partial\vecop\{\widetilde H_t(\bm\theta)\}}/
{\partial\bm\theta\trans}.
$ Then, the sample counterparts of $\mathcal J_0$ and $\mathcal I_0$ are 
\begin{align}
\widehat{\mathcal J}_T
&=
\frac{1}{T}\sum_{t=1}^T
\widetilde{\mathcal D}_t(\widehat{\bm\theta}_T)\trans
\bigl\{
\widetilde H_t(\widehat{\bm\theta}_T)^{-1}
\kron
\widetilde H_t(\widehat{\bm\theta}_T)^{-1}
\bigr\}
\widetilde{\mathcal D}_t(\widehat{\bm\theta}_T),
\label{eq:Jhat}\\
\widehat{\mathcal I}_T
&=
\frac{1}{T}\sum_{t=1}^T
\frac{\partial\widetilde\ell_t(\widehat{\bm\theta}_T)}
{\partial\bm\theta}
\frac{\partial\widetilde\ell_t(\widehat{\bm\theta}_T)}
{\partial\bm\theta\trans}.
\label{eq:Ihat}
\end{align}

\begin{corollary}
\label{lem:avar-est}
Under the conditions of Theorem~\ref{th:an}, if
$\E\|\bm x_t\|^8<\infty$, then
$\widehat{\mathcal J}_T\xrightarrow{p}\mathcal J_0,$ and $
\widehat{\mathcal I}_T\xrightarrow{p}\mathcal I_0.$
\end{corollary}
As a direct consequence of Corollary~\ref{lem:avar-est}, $\widehat{\mathcal V}_T
=
\widehat{\mathcal J}_T^{-1}
\widehat{\mathcal I}_T
\widehat{\mathcal J}_T^{-1}$ yields a consistent estimator of the asymptotic
covariance matrix and thereby justifies large-sample confidence intervals and
score inference for the T-BEKK parameters.

\section{Mode-Specific Volatility Spillover Inference}
\label{sec:spillover}
The mode-specific Kronecker factors allow volatility transmission to be
studied separately along each tensor mode. An off-diagonal entry of \(A_k\)
captures transmission of return shocks across coordinates along mode \(k\),
whereas an off-diagonal entry of \(B_k\) captures transmission through past
conditional covariance. The two factors therefore distinguish short-run ARCH
spillovers from persistent GARCH spillovers.

To summarize the magnitude of these spillover intensities, define
\begin{equation}
\pi_k^A=\frac{\|\offdiag(A_k)\|_F^2}{\|A_k\|_F^2},
\qquad
\pi_k^B=\frac{\|\offdiag(B_k)\|_F^2}{\|B_k\|_F^2},
\label{eq:spillover-intensity}
\end{equation}
where $\offdiag(X)$ sets the diagonal of a square matrix $X$ to zero. These ratios are
the shares of the squared Frobenius norms of the corresponding factors
attributable to off-diagonal terms. They lie in \([0,1]\) and are invariant to scalar rescaling of the
Kronecker factors.

Given a mode $k$, consider the following hypothesis testing problems, 
\begin{equation}
H_{0,k}^A:\offdiag(A_k)=0,\qquad
H_{0,k}^B:\offdiag(B_k)=0,\qquad
H_{0,k}^{\rm dyn}:\offdiag(A_k)=\offdiag(B_k)=0.
\label{eq:spillover-nulls}
\end{equation}
The first two hypotheses concern the ARCH and GARCH effects separately. The
joint null rules out dynamic transmission across coordinates along mode \(k\),
while allowing own ARCH effects and own volatility persistence through the
diagonal entries.

The spillover intensities in \eqref{eq:spillover-intensity} are locally
quadratic around the no-spillover null, where their gradients vanish. Consequently, a Wald test formulated directly in terms
of an intensity has a degenerate first-order expansion under the null. We
therefore test the underlying zero restrictions using a score, or Lagrange
multiplier (LM), test.

For mode $k$, define $r_k^A(\bm\theta)=\vecoff(A_k)$,
$r_k^B(\bm\theta)=\vecoff(B_k)$, and let $r_k^{\rm dyn}$ stack these two
vectors. Here, $\vecoff(M)$ stacks the off-diagonal entries of $M$
columnwise, omitting the diagonal entries. Write
$R_k^j=\partial r_k^j(\bm\theta)/\partial\bm\theta^\top$ and
$\Theta_k^{j,0}=\{\bm\theta:r_k^j(\bm\theta)=0\}$,
$j\in\{A,B,\mathrm{dyn}\}$. Define the sample score by
\[
\widehat{\bm s}_T(\bm\theta)
=T^{-1}\sum_{t=1}^T\nabla_{\bm\theta}\widetilde\ell_t(\bm\theta),
\]
and denote by $\widehat{\bm\theta}_{k,T}^{j,0}$ the restricted QMLE over
$\Theta_k^{j,0}$. Let $\widehat{\mathcal J}_{k,T}^j$ and
$\widehat{\mathcal I}_{k,T}^j$ be the analogues of
\eqref{eq:Jhat} and \eqref{eq:Ihat}, respectively, evaluated at this
restricted estimator, and set
\[
\widehat{\mathcal V}_{k,T}^j
=(\widehat{\mathcal J}_{k,T}^j)^{-1}
\widehat{\mathcal I}_{k,T}^j
(\widehat{\mathcal J}_{k,T}^j)^{-1}.
\]
Define
\begin{equation}
\begin{aligned}
LM_{k,T}^j
={}&T\,\widehat{\bm s}_T^\top
(\widehat{\mathcal J}_{k,T}^j)^{-1}(R_k^j)^\top
\{R_k^j\widehat{\mathcal V}_{k,T}^j(R_k^j)^\top\}^{-1}
R_k^j(\widehat{\mathcal J}_{k,T}^j)^{-1}\widehat{\bm s}_T,
\end{aligned}
\label{eq:spillover-score-stat}
\end{equation}
where the score $\widehat{\bm s}_T$ in \eqref{eq:spillover-score-stat} is evaluated at
$\widehat{\bm\theta}_{k,T}^{j,0}$.

\begin{theorem}
\label{th:spillover-score}
Fix mode $k$ and $j\in\{A,B,\mathrm{dyn}\}$. Suppose that the conditions of
Corollary~\ref{lem:avar-est} hold on the full parameter space $\Theta$ and on the
restricted space $\Theta_k^{j,0}$, with $\bm\theta_0$ in its relative interior.
Suppose further that
$R_k^j\mathcal V_0(R_k^j)^\top$ is nonsingular, where
$\mathcal V_0=\mathcal J_0^{-1}\mathcal I_0\mathcal J_0^{-1}$. Then, under $H_{0,k}^j$,
\[
LM_{k,T}^j\xrightarrow{d}\chi^2_{q_j},
\qquad
q_A=q_B=d_k(d_k-1),\quad q_{\mathrm{dyn}}=2d_k(d_k-1).
\]
\end{theorem}

Thus the null hypotheses in \eqref{eq:spillover-nulls} can be tested by
comparing $LM_{k,T}^j$ with the corresponding chi-square critical value. Away from the no-spillover null, the usual delta method can be used to
conduct inference on spillover magnitude.

\begingroup
\emergencystretch=1em
Under the alternative hypotheses, spillover magnitudes can be summarized by the
intensities $g_k^A(\bm\theta)=\normF{\offdiag(A_k)}^2/\normF{A_k}^2$ and
$g_k^B(\bm\theta)=\normF{\offdiag(B_k)}^2/\normF{B_k}^2$.
Thus, $\pi_k^j=g_k^j(\bm\theta_0)$ and its estimator is
$\widehat\pi_k^j=g_k^j(\widehat{\bm\theta}_T)$ for $j\in\{A,B\}$. For
$g(M)=\normF{\offdiag(M)}^2/\normF{M}^2$, write
\par\endgroup
\begin{equation}
\nabla_M g
=
\frac{
2\{\normF{M}^2\offdiag(M)-\normF{\offdiag(M)}^2M\}
}{\normF{M}^4}.
\label{eq:spillover-gradg}
\end{equation}
where the gradients of $g_k^A$ and $g_k^B$ with respect to $\bm\theta$ are
obtained by applying the chain rule to the corresponding factor. Define
\[
\sigma_{k,A}^2
=
\nabla g_k^A(\bm\theta_0)^\top
\mathcal V_0
\nabla g_k^A(\bm\theta_0),
\qquad
\widehat\sigma_{k,A}^2
=
\nabla g_k^A(\widehat{\bm\theta}_T)^\top
\widehat{\mathcal V}_T
\nabla g_k^A(\widehat{\bm\theta}_T),
\]
and define $\sigma_{k,B}^2$ and $\widehat\sigma_{k,B}^2$ analogously with
$g_k^A$ replaced by $g_k^B$.

\begin{theorem}
\label{th:spillover-intensity}
Suppose that the conditions of Corollary~\ref{lem:avar-est} hold. If
$\pi_k^A>0$ and
$\sigma_{k,A}^2>0$, then
\[
\sqrt T\{g_k^A(\widehat{\bm\theta}_T)-g_k^A(\bm\theta_0)\}
\xrightarrow{d}
\mathcal N(0,\sigma_{k,A}^2),
\qquad
\widehat\sigma_{k,A}^2\xrightarrow{p}\sigma_{k,A}^2.
\]
If $\pi_k^B>0$ and
$\sigma_{k,B}^2>0$, then
\[
\sqrt T\{g_k^B(\widehat{\bm\theta}_T)-g_k^B(\bm\theta_0)\}
\xrightarrow{d}
\mathcal N(0,\sigma_{k,B}^2),
\qquad
\widehat\sigma_{k,B}^2\xrightarrow{p}\sigma_{k,B}^2.
\]
\end{theorem}
The condition $\pi_k^A>0$  (or $\pi_k^B>0$ ) excludes the zero-spillover boundary, at which the
first-order delta method is degenerate. Away from this boundary, we can construct
the confidence interval for spillover intensity as 
$g_k^A(\widehat{\bm\theta}_T)\pm
z_{1-\alpha/2}\widehat\sigma_{k,A}/\sqrt T$, and similarly for $B_k$.  Here, $z_{1-\alpha/2}$ denotes the
$(1-\alpha/2)$th quantile for the standard Gaussian distribution.

The restricted score tests therefore address the absence of mode-specific
spillover, while the delta method intervals describe nonzero spillover
intensities. The next section moves from directional spillover inference to a
portmanteau diagnostic for serial dependence in the quadratic form residuals.

\section{Model checking}
\label{sec:checking}
This section develops a portmanteau diagnostic test for serial dependence in
the quadratic form residuals of the fitted T-BEKK model. Under correct model
specification, we have that
\[
s_t(\bm\theta_0):=\bm x_t^\top H_t(\bm\theta_0)^{-1}\bm x_t
=\bm\eta_t^\top\bm\eta_t.
\]
Hence $u_t(\bm\theta_0):=s_t(\bm\theta_0)-d$ is i.i.d.\ with mean zero. This
implies that its autocorrelations vanish at all nonzero lags. We therefore
construct a portmanteau test from
\(\widehat u_t=s_t(\widehat{\bm\theta}_T)-d\). The test is analogous to squared residual diagnostics for univariate GARCH models in \citet{mcleodli1983} and \citet{lingli1997},  and to related specification tests for conditional
heteroskedasticity in   \citet{limak1994} and \citet{lundberghterasvirta2002}.
Here $s_t(\widehat{\bm\theta}_T)$ is the squared Mahalanobis norm of
$\bm x_t$ under the fitted conditional covariance matrix, so
$\widehat u_t$ summarizes the magnitude of the fitted multivariate innovation
after covariance standardization.

For a lag order
$\ell\ge 1$, define
\begin{equation}
S_\ell=\frac{1}{T}\sum_{t=\ell+1}^T \widehat u_t\widehat u_{t-\ell},
\qquad
\widehat\kappa=\frac{1}{T}\sum_{t=1}^T \widehat u_t^2,
\qquad
\widehat\rho(\ell)=\frac{S_\ell}{\widehat\kappa},
\label{eq:rhohat}
\end{equation}
and collect $\widehat{\bm\rho}_h=\bigl(\widehat\rho(1),\dots,\widehat\rho(h)\bigr)\trans$.

Because \(\bm\theta_0\) is replaced by the QMLE, parameter estimation
contributes to the limiting covariance of the residual autocorrelations. Let
\[
\kappa=\E\bigl[(\bm\eta_t\trans\bm\eta_t-d)^2\bigr],
\qquad
\bm g_t(\bm\theta_0)=
\E\!\left[
\frac{\partial s_t(\bm\theta_0)}{\partial\bm\theta}\Bigm|\mathcal F_{t-1}
\right].
\]
where
\[
g_{t,j}(\bm\theta_0)=
-\Tr\!\left\{
H_t(\bm\theta_0)^{-1}
\frac{\partial H_t(\bm\theta_0)}{\partial\theta_j}
\right\},
\qquad
j=1,\dots,p.
\]
For $\ell=1,\dots,h$, define the $1\times p$ row vectors
\[
M_\ell=
-\E\!\left\{\bm g_t(\bm\theta_0)\trans u_{t-\ell}(\bm\theta_0)\right\},
\qquad
N_\ell=
2\E\!\left[
u_t(\bm\theta_0)u_{t-\ell}(\bm\theta_0)
\frac{\partial\ell_t(\bm\theta_0)}{\partial\bm\theta\trans}
\right],
\]
and let $\bm M_h=(M_1\trans,\dots,M_h\trans)\trans$ and
$\bm N_h=(N_1\trans,\dots,N_h\trans)\trans$. Set
\begin{equation}
\Psi_h=
\kappa^{-2}\Bigl(
\kappa^2 I_h
+\frac{\bm N_h\mathcal J_0^{-1}\bm M_h\trans+\bm M_h\mathcal J_0^{-1}\bm N_h\trans}{2}
+\bm M_h\mathcal J_0^{-1}\mathcal I_0\mathcal J_0^{-1}\bm M_h\trans
\Bigr).
\label{eq:Psih}
\end{equation}

\begin{theorem}
\label{th:portmanteau}
Let $h\ge1$ be fixed. Suppose that the conditions of
Theorem~\ref{th:an} hold, $\E\|\bm x_t\|^8<\infty$,
$\kappa=\Var(\bm\eta_t^\top\bm\eta_t)>0$,
and the T-BEKK model is correctly specified.
Then
\begin{equation}
\sqrt T\,\widehat{\bm\rho}_h\xrightarrow{d}\mathcal N(\bm 0,\Psi_h).
\label{eq:rhoclt}
\end{equation}
If, in addition, $\Psi_h$ is positive definite and $\widehat\Psi_h$ is a
consistent estimator of $\Psi_h$, then
\begin{equation}
Q_T(h)=T\,\widehat{\bm\rho}_h\trans\widehat\Psi_h^{-1}\widehat{\bm\rho}_h\xrightarrow{d}\chi_h^2,
\label{eq:portstat}
\end{equation}
The same asymptotic normality and chi-square limit hold when
$H_t(\widehat{\bm\theta}_T)$ is replaced by $\widetilde H_t(\widehat{\bm\theta}_T)$ in
the residual autocorrelations and the resulting portmanteau statistic.
\end{theorem}

For implementation, let 
$
\widetilde u_t
=
\bm x_t^\top\widetilde H_t(\widehat{\bm\theta}_T)^{-1}\bm x_t-d,
$ $
\widetilde\kappa
=
\sum_{t=1}^T\widetilde u_t^2/T,
$
and 
\[
\widehat g_{t,j}
=-\Tr\!\left\{
\widetilde H_t(\widehat{\bm\theta}_T)^{-1}
\frac{\partial\widetilde H_t(\widehat{\bm\theta}_T)}{\partial\theta_j}
\right\},
\]
and define, for $\ell=1,\dots,h$,
\[
\widehat M_\ell
=-\frac{1}{T}\sum_{t=\ell+1}^T
\widehat{\bm g}_t^\top\widetilde u_{t-\ell},
\qquad
\widehat N_\ell
=\frac{2}{T}\sum_{t=\ell+1}^T
\widetilde u_t\widetilde u_{t-\ell}
\frac{\partial\widetilde\ell_t(\widehat{\bm\theta}_T)}
{\partial\bm\theta^\top}.
\]
After stacking these rows into $\widehat{\bm M}_h$ and
$\widehat{\bm N}_h$, respectively, we obtain $\widehat\Psi_h$ from
\eqref{eq:Psih} by replacing
$(\kappa,\bm M_h,\bm N_h,\mathcal J_0,\mathcal I_0)$ with
$(\widetilde\kappa,\widehat{\bm M}_h,\widehat{\bm N}_h,
\widehat{\mathcal J}_T,\widehat{\mathcal I}_T)$.

The null hypothesis of correct specification is rejected at level $\alpha$
whenever $Q_T(h)>\chi^2_{h,1-\alpha}$, where $\chi^2_{h,1-\alpha}$ is the
$(1-\alpha)$th quantile of the chi-square distribution with $h$ degrees of
freedom. Theorem~\ref{th:portmanteau}
is stated for fixed $h$. In practice, the choice of the fixed lag truncation $h$ depends on the
sampling frequency and finite-sample performance. Moderate fixed values such
as $h=6$ or $h=8$ are typically adequate for the sample sizes considered here;
see e.g. \citet{tsay2008}. A rejection indicates remaining serial dependence in the
quadratic form residuals and  provides evidence against the adequacy
of the fitted conditional covariance specification.

\section{A factor extension: TF-BEKK}
\label{sec:factor}

When one or more mode dimensions are large relative to the available sample
size, a low-dimensional factor representation reduces the dimension of the
covariance matrices used in likelihood evaluation. We therefore consider a low-rank extension in which the observed
tensor is driven by a lower-dimensional latent tensor whose conditional
covariance follows a T-BEKK model. This extension links the model to
the literature on factor-based covariance reduction and recent tensor factor methods
\citep{lanne2007,engleledoitwolf2019,chen2022tensor,chen2024preavg,
chen2024semiparametric,han2024cp,barigozzi2025tailrobust} while retaining a
T-BEKK specification for latent volatility.
\subsection{TF-BEKK model}
We consider the following Tucker factor representation
\begin{equation}
\tensor X_t = \tensor S_t^*\tensormult{1}{R_1^*}\cdots\tensormult{m}{R_m^*}+\tensor E_t,
\label{eq:rawfactor}
\end{equation}
where $\tensor S_t^*\in\R^{r_1\times\cdots\times r_m}$ with $r_k\ll d_k$
is the latent common factor tensor, $R_k^*\in\R^{d_k\times r_k}$ satisfying
$(R_k^*)^\top R_k^*/d_k=I_{r_k}$ is the mode-$k$ loading matrix, and
$\tensor E_t\in\R^{d_1\times\cdots\times d_m}$ is the idiosyncratic noise
tensor. Define $U_k=R_k^*/\sqrt{d_k}$ and $P_k=U_kU_k^\top$. Thus
$U_k^\top U_k=I_{r_k}$, and $P_k$ is the orthogonal projector onto the
population mode-$k$ loading space.

As in conventional factor models, $P_k$ is identified, whereas the particular
basis $U_k$ is identified only up to an orthogonal rotation. Indeed, a
rotation of $U_k$ can be offset by the corresponding inverse rotation of the
factor tensor without changing the distribution of $\tensor X_t$. Because
the parameters governing the factor volatility depend on the chosen factor
coordinates, we fix an orientation before specifying the latent T-BEKK model.

For each mode $k$, let $J_k\in\R^{d_k\times r_k}$ be a prespecified reference matrix
satisfying Assumption~S.2.4 in the supplement. Define
\begin{equation}
U_k^\circ
=P_kJ_k(J_k^\top P_kJ_k)^{-1/2},
\qquad
R_k^\circ=\sqrt{d_k}\,U_k^\circ.
\label{eq:factor-anchor-orientation}
\end{equation}
Both $U_k$ and $U_k^\circ$ are orthonormal bases for
$\operatorname{range}(P_k)$, so $U_k^\circ=U_kG_{k,0}$ for an orthogonal
matrix $G_{k,0}\in\mathcal O(r_k)$. The corresponding oriented factor tensor is
\[
\tensor S_t^\circ
=
\tensor S_t^*
\tensormult{1}{G_{1,0}^\top}
\cdots
\tensormult{m}{G_{m,0}^\top}.
\]
The factor representation in \eqref{eq:rawfactor} can then be rewritten as
\begin{equation}
\tensor X_t
=
\tensor S_t^\circ
\tensormult{1}{R_1^\circ}
\cdots
\tensormult{m}{R_m^\circ}
+\tensor E_t.
\label{eq:factor-model-main}
\end{equation}

Let $\bm s_t^\circ=\vecop(\tensor S_t^\circ)$ and
$r=\prod_{k=1}^m r_k$. Let
$\mathcal G_t=\sigma(\bm s_u^\circ,\bm e_u:u\le t)$ denote the joint
filtration generated by information up to time   $t$, where $\bm e_t=\vecop(\tensor E_t)$. We write the
conditional volatility representation of the oriented factor process
$\{\bm s_t^\circ\}$ by assuming a T-BEKK model as
\begin{flalign}
    \bm s_t^\circ
=&
(H_{f,t}^\circ)^{1/2}\bm\eta_{f,t}^\circ,\label{eq:factor}\\
H_{f,t}^\circ
=&
\Omega_f^\circ+A_f^\circ\bm s_{t-1}^\circ(\bm s_{t-1}^\circ)\trans (A_f^\circ)\trans
+ B_f^\circ H_{f,t-1}^\circ(B_f^\circ)\trans,\label{eq:TF-BEKK}
\end{flalign}
Here $\{\bm\eta_{f,t}^\circ\}$ is a sequence of $r$-dimensional i.i.d.\ innovations  such that $\bm\eta_{f,t}^\circ$ is independent of
$\mathcal G_{t-1}$ and satisfies $\E(\bm\eta_{f,t}^\circ)=\bm 0$ and
$\E\{\bm\eta_{f,t}^\circ(\bm\eta_{f,t}^\circ)^\top\}=I_r$. Moreover, \[
\Omega_f^\circ=\Omega_{f,m}^\circ\kron\cdots\kron\Omega_{f,1}^\circ,
\quad
A_f^\circ=A_{f,m}^\circ\kron\cdots\kron A_{f,1}^\circ,
\quad
B_f^\circ=B_{f,m}^\circ\kron\cdots\kron B_{f,1}^\circ.
\]
Let $\bm\theta_f$ collect the free parameters in the modewise matrices above
under the corresponding Kronecker normalizations, let $\Theta_f$ denote its
parameter space, and let $\bm\theta_{f,0}$ denote its true value for the
oriented factor process.

We refer to \eqref{eq:factor-model-main}--\eqref{eq:TF-BEKK} as the
tensor factor-BEKK (TF-BEKK) model.

Writing
$\mathcal R^\circ=R_m^\circ\kron\cdots\kron R_1^\circ$, we have
$\bm x_t=\mathcal R^\circ\bm s_t^\circ+\bm e_t$.
If
$\E(\bm e_t\mid\mathcal G_{t-1})=\bm 0$,
$\Sigma_e:=\E(\bm e_t\bm e_t^\top\mid\mathcal G_{t-1})$ is time-invariant, and
$\E\{\bm e_t(\bm s_t^\circ)^\top\mid\mathcal G_{t-1}\}=\bm 0$, then the
structural conditional covariance of the observed tensor given
$\mathcal G_{t-1}$ is
\[
\Sigma_{x,t}:=\E(\bm x_t\bm x_t^\top\mid\mathcal G_{t-1})
=
\mathcal R^\circ H_{f,t}^\circ(\bm\theta_{f,0})(\mathcal R^\circ)^\top+\Sigma_e.
\]
 
Under this specification, the time-varying component of the observed
conditional covariance is generated by the low-dimensional factor recursion,
whereas the idiosyncratic covariance is time invariant.

\subsection{Two-step estimation}
The TF-BEKK model is estimated in two steps. The first step estimates the
mode-specific loading spaces and recovers the latent factor tensors, whereas
the second step estimates the parameters governing the conditional covariance
of the recovered latent factors.

For each mode $k$, let $d_{-k}=d/d_k$ and let $X_{t,(k)}$ denote the mode-$k$
unfolding of $\tensor X_t$. We estimate the loading space from the
contemporaneous modewise second-order moment
\begin{equation}
\widehat M_k
=
\frac{1}{T d_{-k}}
\sum_{t=1}^T X_{t,(k)}X_{t,(k)}^\top,
\qquad
\widehat P_k    
=
\operatorname{Proj}_{r_k}(\widehat M_k),
\label{eq:modewise-pca}
\end{equation}
where $\operatorname{Proj}_{r_k}(M)$ denotes the projector onto the leading
$r_k$ eigenspace of a symmetric matrix $M$. We refer to $\widehat P_k$ as the
modewise principal component analysis (modewise PCA) estimator of the
mode-$k$ loading space. Equivalently, it is obtained from the leading
$r_k$ left singular vectors of the matrix formed by concatenating the
mode-$k$ unfoldings over time. This construction is the modewise PCA
estimator studied by \citet{zhang2026tucker}; see also the tensor PCA and
contemporaneous PCA estimators of \citet{babii2025tpca} and
\citet{barigozzi2026statistical}, respectively.
 
\begin{remark}
The TF-BEKK factor process is a martingale difference sequence, so its
nonzero-lag autocovariances vanish. Hence, autocovariance-based estimators and
their iterative refinements \citep{chen2022tensor,han2024iterative} are not
directly applicable under the present specification. We use the
contemporaneous modewise PCA estimator, which yields the first-step rates
established below.
\end{remark}

Using the same reference matrix $J_k$, define the sample orientation by
\begin{equation}
\widehat U_k^\circ
=\widehat P_kJ_k(J_k^\top\widehat P_kJ_k)^{-1/2},
\qquad
\widehat R_k^\circ=\sqrt{d_k}\,\widehat U_k^\circ.
\label{eq:sample-anchor-orientation}
\end{equation}
Under Assumption~S.2.4, the inverse is well defined with
probability tending to one. The oriented factor tensors are then estimated by
the normalized projection
\begin{equation}
\begin{aligned}
\widehat{\tensor S}_t^\circ
 =\tensor X_t
\tensormult{1}{\frac{(\widehat R_1^\circ)^\top}{d_1}}
\cdots
\tensormult{m}{\frac{(\widehat R_m^\circ)^\top}{d_m}}=
\tensor X_t
\tensormult{1}{\frac{(\widehat U_1^\circ)^\top}{\sqrt{d_1}}}
\cdots
\tensormult{m}{\frac{(\widehat U_m^\circ)^\top}{\sqrt{d_m}}}.
\end{aligned}
\label{eq:factor-estimator}
\end{equation}

We then provide the convergence rate of the first-step estimation under conditions similar to those in
\citet[Theorem~1]{barigozzi2026statistical}; see Assumption~S.2.1 in the supplement for details.
\begin{proposition}
\label{prop:factorrate}
Suppose Assumption~S.2.1 in the supplement holds, and let
\begin{equation}
\delta_T
=
\max_{1\le k\le m}
\left\{(Td_{-k})^{-1/2}+d_k^{-1}\right\}.
\label{eq:factor-delta-rate}
\end{equation}
Then, 
\begin{equation}
\max_{1\le k\le m}\|\widehat P_k-P_k\|_S
=O_p(\delta_T),
\qquad
\max_{1\leq k\leq m}
\|\widehat U_k^\circ-U_k^\circ\|_F
=
O_p(\delta_T).
\label{eq:first-step-rate}
\end{equation}
Moreover,
\begin{equation}
\frac{1}{T}\sum_{t=1}^T
\|\widehat{\tensor S}_t^\circ-\tensor S_t^\circ\|_F^2
=
O_p\!\left(\delta_T^2+d^{-1}\right),
\label{eq:factorrateavg}
\end{equation}
and, for each fixed $t$,
$
\|\widehat{\tensor S}_t^\circ-\tensor S_t^\circ\|_F
=
O_p\!\left(\delta_T+d^{-1/2}\right).
$
\end{proposition}

In the second step, we use the recovered factor series as inputs to the latent
T-BEKK likelihood. Specifically, let
$\widehat{\bm s}_t^\circ=\vecop(\widehat{\tensor S}_t^\circ)$ and estimate
the latent volatility parameter $\bm\theta_f$ by minimizing the negative Gaussian
quasi-log-likelihood constructed from
$\{\widehat{\bm s}_t^\circ\}_{t=1}^T$ given by 
\begin{equation}
\widehat L_{f,T}(\bm\theta_f)=\frac{1}{T}\sum_{t=1}^T \widehat\ell_{f,t}(\bm\theta_f),
\qquad
\widehat\ell_{f,t}(\bm\theta_f)
=
\log\det \widehat H_{f,t}^\circ(\bm\theta_f)
+
(\widehat{\bm s}_t^\circ)\trans\widehat H_{f,t}^\circ(\bm\theta_f)^{-1}\widehat{\bm s}_t^\circ.
\label{eq:feasible-factor-like}
\end{equation}
Here $\widehat H_{f,t}^\circ(\bm\theta_f)$ follows the recursion in
\eqref{eq:TF-BEKK} with $\widehat{\bm s}_{t-1}^\circ$ in place of
$\bm s_{t-1}^\circ$.  The two-step QMLE is
$\widehat{\bm\theta}_{f,T}
=\operatorname*{arg\,min}_{\bm\theta_f\in\Theta_f}
\widehat L_{f,T}(\bm\theta_f)$.

Let \(\bm\theta_{f,0}\) be the oracle parameter for the oriented factor
process \(\{\bm s_t^\circ\}\). Note we assume that the ranks
\(r_1,\ldots,r_m\), the latent dimension \(r=\prod_{k=1}^m r_k\), and the
dimension of \(\bm\theta_f\) are fixed. Let
\(\mathcal J_f\) and \(\mathcal I_f\) be defined analogously to  
\(\mathcal J_0\) and \(\mathcal I_0\), respectively.

\begin{theorem}
\label{cor:factortransfer}
Suppose the conditions of Proposition~\ref{prop:factorrate} and
Assumptions~S.2.6 and~S.2.7 in the
supplement hold.
\begin{enumerate}[label=\upshape(\roman*)]
    \item If $\delta_T^2+d^{-1}\to0$ and the conditions in
    Theorem~\ref{th:consistency} hold for the corresponding rotated oracle
    T-BEKK problem based on $\{\tensor S_t^\circ\}$, then
    $\widehat{\bm\theta}_{f,T}\xrightarrow{p}\bm\theta_{f,0}$.
    \item If, in addition, $\sqrt T(\delta_T+d^{-1/2})\to0$ and the
    conditions in Theorem~\ref{th:an} hold for the corresponding rotated
    oracle T-BEKK problem, then
    \begin{equation*}
    \sqrt T\,(\widehat{\bm\theta}_{f,T}-\bm\theta_{f,0})
    \xrightarrow{d}
    \mathcal N\!\bigl(\bm 0,\mathcal J_f^{-1}\mathcal I_f\mathcal J_f^{-1}\bigr).
    \end{equation*}
\end{enumerate}
\end{theorem}

 Theorem~\ref{cor:factortransfer} (i)
establishes consistency under consistent factor recovery, and
part~\textup{(ii)} gives oracle equivalence when
$\sqrt T(\delta_T+d^{-1/2})\to0$; see similar assumptions in  \cite{yu2025matrix}. A plug-in result for the reconstructed covariance
matrix $\Sigma_{x,t}$ is given in
Proposition~S.2.10 in the supplement. The resulting covariance estimator can be used in subsequent decision problems, such as the portfolio allocation  in Section~\ref{sec:application2}.

\section{Simulations}
\label{sec:simulation}
\noindent This section  conducts three simulation experiments to examine the finite-sample
performance of the QMLE, the mode-specific spillover tests, and the
portmanteau test.
\subsection{Finite-sample properties of the T-BEKK QMLE}
We first examine the finite-sample performance of the Gaussian QMLE of the T-BEKK model. For each sample size
$T\in\{1000,2000,4000\}$ and each innovation distribution, we generate 1,000
Monte Carlo replications of a $2\times2$ matrix-valued process $\tensor X_t$.
Let $\bm x_t=\vecop(\tensor X_t)\in\mathbb{R}^4$. The process is generated from
the T-BEKK$(1,1)$ model
	\begin{equation}
	\begin{aligned}
	\bm x_t
	&=
	H_t^{1/2}\bm\eta_t,
	\\
	H_t
	&=
	\Omega
	+
	(A_2\kron A_1)\bm x_{t-1}\bm x_{t-1}^{\top}(A_2\kron A_1)^{\top} 
	+
	(B_2\kron B_1)H_{t-1}(B_2\kron B_1)^{\top},
	\end{aligned}
	\label{eq:sim_tbekk_dgp}
	\end{equation}
where
	$
	\Omega=(C_2C_2^\top)\kron(C_1C_1^\top).
	$ Here, we consider two innovation distributions for $\bm\eta_t$: standard Gaussian
innovations and innovations with independent standardized Student's $t_{10}$ components.
Both innovation distributions have mean zero, covariance matrix $I_4$, and
finite eighth moments.
The parameter matrices are given by
\begingroup
\small
\setlength{\arraycolsep}{3pt}
\renewcommand{\arraystretch}{0.9}
\[
\begin{array}{@{}r@{\;=\;}r@{\hspace{1.5em}}r@{\;=\;}r@{\hspace{1.5em}}r@{\;=\;}r@{}}
C_1 &
\begin{pmatrix}
0.7 & 0\\
0.5 & 0.4
\end{pmatrix},
&
A_1 &
\begin{pmatrix}
0.4 & -0.1\\
0.2 & 0.4
\end{pmatrix},
&
B_1 &
\begin{pmatrix}
0.5 & 0.2\\
-0.15 & 0.6
\end{pmatrix},
\\[0.8ex]
C_2 &
\begin{pmatrix}
1 & 0\\
0.4 & 0.8
\end{pmatrix},
&
A_2 &
\begin{pmatrix}
1 & 0.3\\
-0.3 & 0.7
\end{pmatrix},
&
B_2 &
\begin{pmatrix}
1 & -0.3\\
0.4 & 0.8
\end{pmatrix}.
\end{array}
\]
\endgroup

The $(1,1)$ entries of the mode 2 factors $C_2$, $A_2$, and
$B_2$ are fixed at one under the normalization in Assumption~\ref{ass:id}; hence
they are not reported as free parameters. In addition, for the vectorized
coefficient matrices
$A=A_2\kron A_1$ and $B=B_2\kron B_1$, we numerically verify that
$\rho(A\kron A+B\kron B)<1$, so the data-generating process lies in the
stationary region covered by Proposition~\ref{prop:stationarity}. The
off-diagonal entries in $A_k$ and $B_k$ generate cross entry volatility
transmission through the ARCH and GARCH components. Under this normalization,
T-BEKK$(1,1)$ has 19 free parameters, compared with 42 under unrestricted
BEKK$(1,1)$.

Table~\ref{tab:qmle_main} reports the  bias, root mean
squared error (RMSE), and average estimated standard error
(ASE) for $T=1000$ and $T=2000$. The ASE averages the sandwich standard
errors across Monte Carlo replications. The results of $T=4000$ are reported in
Table~S.1 and~S.2 of the supplement.

\begin{table}[!htbp]
\centering
\caption{QMLE finite-sample performance}
\label{tab:qmle_main}
\footnotesize
\renewcommand{\arraystretch}{1.02}
\setlength{\tabcolsep}{2.8pt}
\begin{tabular}{@{}l*{12}{c}@{}}
\toprule
\multirow{3}{*}{Parameter}
& \multicolumn{6}{c}{Gaussian}
& \multicolumn{6}{c}{Standardized Student's $t_{10}$} \\
\cmidrule(lr){2-7}\cmidrule(l){8-13}
& \multicolumn{3}{c}{$T=1000$}
& \multicolumn{3}{c}{$T=2000$}
& \multicolumn{3}{c}{$T=1000$}
& \multicolumn{3}{c}{$T=2000$} \\
\cmidrule(lr){2-4}\cmidrule(lr){5-7}\cmidrule(lr){8-10}\cmidrule(l){11-13}
& Bias & RMSE & ASE & Bias & RMSE & ASE
& Bias & RMSE & ASE & Bias & RMSE & ASE \\
\midrule
$(C_1)_{11}$ & 0.004 & 0.069 & 0.055 & 0.004 & 0.045 & 0.041 & 0.003 & 0.071 & 0.058 & 0.005 & 0.050 & 0.043 \\
$(C_1)_{21}$ & -0.009 & 0.065 & 0.046 & -0.004 & 0.041 & 0.035 & -0.010 & 0.066 & 0.048 & -0.006 & 0.044 & 0.036 \\
$(C_1)_{22}$ & -0.006 & 0.051 & 0.035 & -0.001 & 0.028 & 0.026 & -0.007 & 0.055 & 0.037 & -0.001 & 0.032 & 0.028 \\
\addlinespace[0.2ex]
$(A_1)_{11}$ & 0.001 & 0.045 & 0.042 & 0.002 & 0.032 & 0.030 & 0.002 & 0.047 & 0.045 & 0.002 & 0.033 & 0.033 \\
$(A_1)_{21}$ & 0.001 & 0.039 & 0.036 & 0.002 & 0.027 & 0.025 & 0.004 & 0.042 & 0.039 & 0.003 & 0.029 & 0.028 \\
$(A_1)_{12}$ & -0.001 & 0.039 & 0.036 & -0.001 & 0.027 & 0.026 & -0.002 & 0.040 & 0.037 & -0.002 & 0.027 & 0.027 \\
$(A_1)_{22}$ & -0.002 & 0.041 & 0.039 & -0.001 & 0.028 & 0.028 & -0.003 & 0.044 & 0.043 & -0.002 & 0.030 & 0.030 \\
\addlinespace[0.2ex]
$(B_1)_{11}$ & -0.037 & 0.149 & 0.117 & -0.022 & 0.099 & 0.088 & -0.035 & 0.146 & 0.122 & -0.025 & 0.103 & 0.092 \\
$(B_1)_{21}$ & 0.002 & 0.135 & 0.086 & 0.000 & 0.081 & 0.062 & 0.005 & 0.125 & 0.093 & -0.001 & 0.085 & 0.067 \\
$(B_1)_{12}$ & 0.010 & 0.098 & 0.081 & 0.005 & 0.064 & 0.059 & 0.006 & 0.101 & 0.086 & 0.005 & 0.067 & 0.062 \\
$(B_1)_{22}$ & -0.016 & 0.095 & 0.076 & -0.008 & 0.058 & 0.053 & -0.019 & 0.100 & 0.082 & -0.009 & 0.065 & 0.057 \\
\addlinespace[0.2ex]
$(C_2)_{21}$ & 0.001 & 0.091 & 0.065 & -0.002 & 0.056 & 0.048 & 0.003 & 0.089 & 0.067 & -0.001 & 0.058 & 0.049 \\
$(C_2)_{22}$ & -0.006 & 0.092 & 0.067 & -0.004 & 0.055 & 0.048 & -0.004 & 0.097 & 0.073 & -0.005 & 0.064 & 0.052 \\
\addlinespace[0.2ex]
$(A_2)_{21}$ & -0.004 & 0.053 & 0.050 & -0.001 & 0.036 & 0.035 & -0.001 & 0.057 & 0.053 & -0.001 & 0.038 & 0.037 \\
$(A_2)_{12}$ & 0.000 & 0.078 & 0.072 & -0.002 & 0.051 & 0.051 & 0.000 & 0.083 & 0.079 & -0.002 & 0.058 & 0.056 \\
$(A_2)_{22}$ & 0.003 & 0.076 & 0.072 & -0.001 & 0.052 & 0.051 & -0.004 & 0.083 & 0.078 & -0.004 & 0.056 & 0.055 \\
\addlinespace[0.2ex]
$(B_2)_{21}$ & 0.011 & 0.090 & 0.080 & 0.007 & 0.061 & 0.056 & 0.011 & 0.099 & 0.083 & 0.006 & 0.063 & 0.057 \\
$(B_2)_{12}$ & 0.006 & 0.179 & 0.132 & 0.009 & 0.102 & 0.089 & 0.004 & 0.176 & 0.139 & 0.012 & 0.108 & 0.094 \\
$(B_2)_{22}$ & 0.021 & 0.201 & 0.159 & 0.018 & 0.119 & 0.109 & 0.027 & 0.216 & 0.170 & 0.026 & 0.142 & 0.118 \\
\bottomrule
\end{tabular}
\end{table}

Table~\ref{tab:qmle_main} yields three main findings. First, the biases are
generally small relative to the corresponding RMSEs. Second, under both
innovation distributions, every reported RMSE declines as $T$ increases from
1000 to 2000, and the gaps between the RMSEs and ASEs generally decrease. Third, the ARCH parameters $A_k$ are estimated more accurately than the GARCH parameters $B_k$; for the latter, the ASEs approach the corresponding RMSEs from below as $T$ increases.
The results for $T=4000$ reported in the supplement confirm the improvement
with increasing $T$.

\subsection{Finite-sample properties of the mode-specific spillover tests}

We next evaluate the empirical size and power of the restricted score tests in
Theorem~\ref{th:spillover-score}. The designs use a $2\times2$ Gaussian
T-BEKK model, with $\bm x_t=\vecop(\tensor X_t)\in\R^4$,
$\bm\eta_t\sim \mathcal N(\bm 0,I_4)$ independently over time, and
\[
\Omega=(C_2C_2^\top)\kron(C_1C_1^\top),
\qquad
A=A_2\kron A_1,\qquad
B=B_2\kron B_1.
\]
The intercept factors are fixed across all designs:
\begingroup
\small
\setlength{\arraycolsep}{3pt}
\renewcommand{\arraystretch}{0.9}
\[
C_1=
\begin{pmatrix}
0.7 & 0\\
0.5 & 0.4
\end{pmatrix},
\qquad
C_2=
\begin{pmatrix}
1 & 0\\
0.4 & 0.8
\end{pmatrix}.
\]
\endgroup
The dynamic factors are chosen to isolate spillover along mode 2. In both
spillover designs, the mode-1 ARCH and GARCH factors are kept diagonal:
\begingroup
\small
\setlength{\arraycolsep}{3pt}
\renewcommand{\arraystretch}{0.9}
\[
A_1=A_1^{(0)}=
\begin{pmatrix}
0.35 & 0\\
0 & 0.3
\end{pmatrix},
\qquad
B_1=B_1^{(0)}=
\begin{pmatrix}
0.45 & 0\\
0 & 0.5
\end{pmatrix},
\]
\endgroup
The two designs differ only in which mode-2 dynamic factor contains the
off-diagonal perturbation.

\noindent\textit{Design S1 (ARCH factor spillover).} The off-diagonal
perturbation is confined to the ARCH factor:
\begingroup
\small
\setlength{\arraycolsep}{3pt}
\renewcommand{\arraystretch}{0.9}
\[
A_2=A_2^{A}(\tau)=
\begin{pmatrix}
1 & \tau\\
\tau & 0.7
\end{pmatrix},
\qquad
B_2=B_2^{(0)}=
\begin{pmatrix}
1 & 0\\
0 & 0.6
\end{pmatrix}.
\]
\endgroup

\begin{figure}[!htbp]
\centering
\includegraphics[width=1\linewidth]{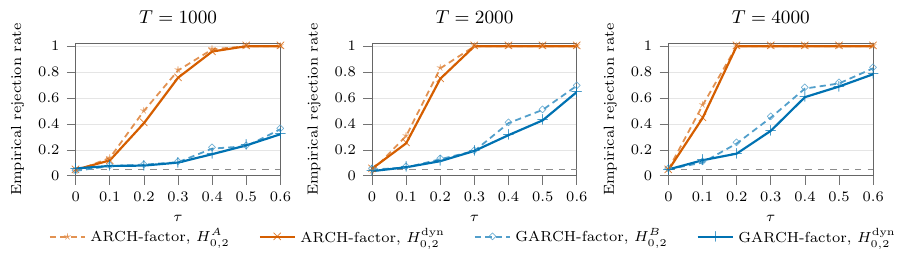}
\caption{Empirical rejection rates of the restricted score tests for mode-2
spillover under Design~S1 (ARCH factor spillover) and Design~S2 (GARCH factor
spillover). The horizontal axis is the perturbation parameter $\tau$, and the
dashed line indicates the nominal $5\%$ level.}
\label{fig:spillover-score-power}
\end{figure}

\noindent\textit{Design S2 (GARCH factor spillover).} The off-diagonal
perturbation is confined to the GARCH factor:
\begingroup
\small
\setlength{\arraycolsep}{3pt}
\renewcommand{\arraystretch}{0.9}
\[
A_2=A_2^{(0)}=
\begin{pmatrix}
1 & 0\\
0 & 0.7
\end{pmatrix},
\qquad
B_2=B_2^{B}(\tau)=
\begin{pmatrix}
1 & \tau\\
\tau & 0.6
\end{pmatrix}.
\]
\endgroup
The scalar $\tau\ge0$ controls the off-diagonal perturbation. We use
$\tau\in\{0,0.1,\ldots,0.6\}$; $\tau=0$ denotes the null of  no-spillover, and positive
values generate mode-2 alternatives.  For each design, $\tau$, and
$T\in\{1000,2000,4000\}$, we use 1,000 Monte Carlo replications and report the
mode-2 factor-specific and joint dynamic tests in
Figure~\ref{fig:spillover-score-power}. 

From Figure~\ref{fig:spillover-score-power}, we find that (i) all tests have
rejection rates close to the nominal \(5\%\) level when \(\tau=0\); (ii) power
increases with both \(\tau\) and \(T\); and (iii) the factor-specific test is
slightly more powerful than the joint test against its corresponding
alternative. The ARCH spillover tests attain power close to one for moderate
values of \(\tau\), whereas the power of the GARCH spillover tests increases more gradually.

	\subsection{Finite-sample properties of the portmanteau test}
	
We next study the finite-sample size and power of the portmanteau statistic
$Q_T(h)$. Under the null, the process follows the correctly specified Gaussian
T-BEKK model in \eqref{eq:sim_tbekk_dgp}. For power, we fit the
T-BEKK$(1,1)$ null model to data generated from the T-BEKK$(2,2)$ model defined as follows
\[
\begin{aligned}
\bm x_t
&=
H_t^{1/2}\bm\eta_t,
\\
H_t
&=
\Omega
+ A \bm x_{t-1} \bm x_{t-1}^\top A^\top
+ B H_{t-1} B^\top
+ A_{\text{extra}} \bm x_{t-2} \bm x_{t-2}^\top A_{\text{extra}}^\top
+ B_{\text{extra}} H_{t-2} B_{\text{extra}}^\top,
\end{aligned}
\]
where $\bm\eta_t\sim\mathcal N(\bm 0,I_4)$ independently over time, and
$\Omega$, $A=A_2\kron A_1$, and $B=B_2\kron B_1$ use the baseline values in
\eqref{eq:sim_tbekk_dgp}. We consider two lag-2 alternatives.

\noindent\textit{Design P1 (lag-2 ARCH).} The departure from the null adds a
lag-2 ARCH term, with
		\[
		A_{\text{extra}} = \delta \|A\|_F D_A, \qquad B_{\text{extra}} = 0,
		\qquad
		D_A = \frac{1}{\sqrt{d}} \operatorname{diag}(1,-1,1,-1).
		\]
Here $d=4$ in the present $2\times2$ design.

\noindent\textit{Design P2 (lag-2 GARCH).} The departure from the null adds a
lag-2 GARCH term, with
		\[
		A_{\text{extra}} = 0, \qquad B_{\text{extra}} = \delta \|B\|_F D_B,
		\]
where $D_B=d^{-1/2}\operatorname{diag}(1,-1,1,-1)$ is defined analogously.
The perturbation parameter $\delta\ge0$ controls the degree of misspecification,
and the Frobenius norm scaling normalizes the added lag-2 coefficient. The
reported grid uses stationary lag-2 alternatives and 1,000 Monte Carlo
replications for each $T \in \{1000,2000,4000\}$ and
$h \in \{4,6,8,10\}$.
	
	\begin{figure}[!htbp]
		\centering
		\includegraphics[width=0.88\linewidth]{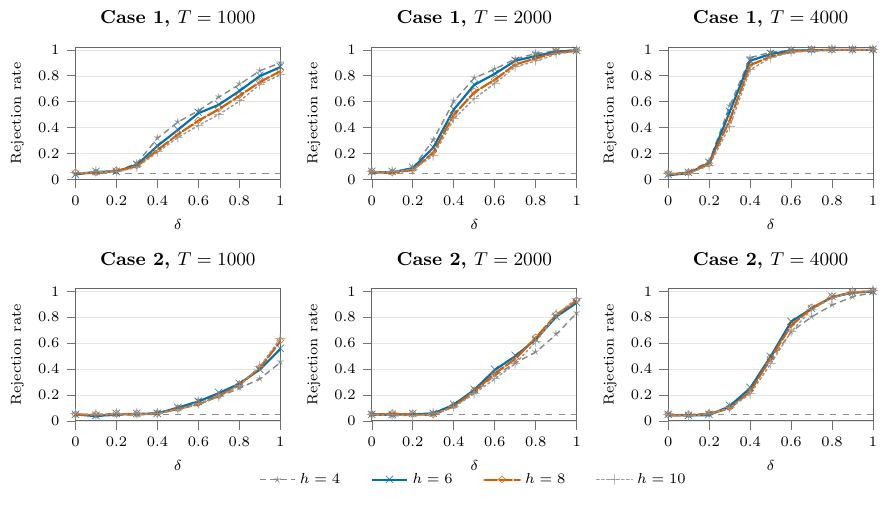}
		\caption{Empirical rejection rates of the portmanteau statistic $Q_T(h)$ under the null and the lag-2 alternatives. Case 1 and Case 2 correspond to Designs~P1 and P2, respectively. The four curves use $h=4,6,8,10$, and the dashed line indicates the nominal 5\% level.}
		\label{fig:power}
	\end{figure}
	
At $\delta=0$, the rejection rates in Figure~\ref{fig:power} are close to the
nominal 5\% level. Under both alternatives, rejection rates increase with
$\delta$ and $T$, with modest differences across
$h\in\{4,6,8,10\}$. The test is more sensitive to the omitted lag-2 ARCH term
than to the omitted lag-2 GARCH term in these designs.

\section{Applications}
\label{sec:application}

We consider two empirical applications involving a matrix of currency futures returns and a tensor of Chinese equity portfolio returns. In the first application, we fit a T-BEKK model to the currency futures matrix and use the fitted model for diagnostic checking and inference on mode-specific volatility spillovers. In the second application, we evaluate conditional covariance forecasts for the Chinese equity tensor through constrained minimum-variance portfolio allocation.
\subsection{Application 1: Currency futures matrix}
\label{sec:application1}

We first consider a \(2\times2\) matrix of weekly currency futures returns, viewed as a tensor with two modes. The row mode distinguishes funding currencies from commodity-linked currencies, while the column mode indexes positions within each currency group. The first row consists of Japanese yen and Swiss franc futures, and the second row consists of Australian dollar and Canadian dollar futures.\footnote{Daily adjusted closing prices are obtained from Yahoo Finance for the CME-listed currency futures with Yahoo Finance symbols \texttt{6J=F}, \texttt{6S=F}, \texttt{6A=F}, and \texttt{6C=F}.} The mode-specific coefficients are interpreted with respect to this fixed arrangement.

For each futures price series, we compute the Friday-to-Friday weekly log return, expressed in percentage points, as
\(
y_t=100\{\log(P_t)-\log(P_{t-1})\},
\)
where \(P_t\) denotes the last available adjusted closing price in each Friday-ending week. We estimate a VAR(5) model for the conditional mean using the 1340 weekly returns in the sample period January 1, 2001, to September 4, 2026. This is the lowest lag order for which all four filtered return series pass the Ljung--Box tests at lags 4, 8, and 12 at the 5\% level. All 1335 resulting residuals, from February 9, 2001, to September 4, 2026, are used for volatility estimation; the minimum \(p\)-value across the twelve mean diagnostics is 0.069. In contrast, the squared residuals exhibit significant serial dependence in three of the four series, providing evidence of conditional heteroskedasticity and motivating the modeling of conditional covariance dynamics. Further details on the lag grid diagnostics are reported in Table~S.3 of the supplement.
\begin{figure}[!htbp]
\centering
\includegraphics[width=0.82\linewidth]{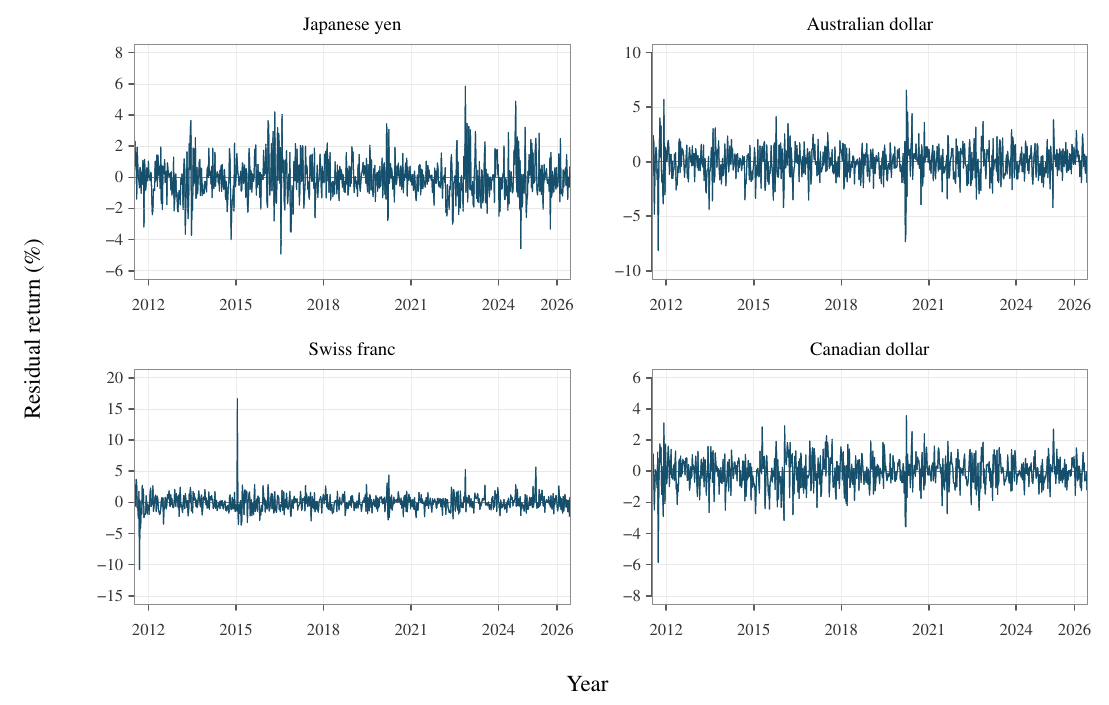}
\caption{Weekly log returns for the currency futures matrix after VAR(5) mean adjustment over the estimation period from 2001-02-09 to 2026-09-04. Each panel displays one entry of the \(2\times2\) matrix.}
\label{fig:app1_fx_series}
\end{figure}

The normalized T-BEKK\((1,1)\) model has 19 free parameters. Estimation
under the stationarity constraint yields
\[
\widehat\rho_{\mathrm{sp}}
=\rho(\widehat A\kron\widehat A+\widehat B\kron\widehat B)
=0.99556.
\]
This spectral radius measures volatility persistence and indicates slowly
decaying volatility shocks.

Table~\ref{tab:app1_qmle} reports the QMLEs and their estimated standard
errors (SEs) for the fitted model.

\begin{table}[htbp]
\centering
\caption{QMLEs and estimated standard errors}
\label{tab:app1_qmle}
\small
\setlength{\tabcolsep}{7pt}
\begin{tabular}{lcccc}
\toprule
Component & \((1,1)\) & \((2,1)\) & \((1,2)\) & \((2,2)\) \\
\midrule
\(\widehat C_1\) & 0.246 (0.034) & 0.046 (0.024) & 0 & 0.069 (0.024) \\
\(\widehat A_1\) & 0.294 (0.041) & 0.062 (0.018) & -0.020 (0.031) & 0.231 (0.044) \\
\(\widehat B_1\) & 0.947 (0.012) & -0.015 (0.005) & 0.001 (0.007) & 0.973 (0.008) \\
\midrule
\(\widehat C_2\) & 1 & 0.686 (0.104) & 0 & 0.994 (0.184) \\
\(\widehat A_2\) & 1 & 0.109 (0.078) & 0.009 (0.075) & 0.787 (0.171) \\
\(\widehat B_2\) & 1 & -0.002 (0.005) & -0.006 (0.006) & 1.002 (0.010) \\
\bottomrule
\end{tabular}
\begin{flushleft}
\footnotesize
Note: Values in parentheses are sandwich SEs. Unit entries are imposed by normalization, and the zero entries in \(C_1\) and \(C_2\) arise from their Cholesky forms.
\end{flushleft}
\end{table}

We next apply the portmanteau diagnostic test developed in Section~\ref{sec:checking} to the residuals of the fitted T-BEKK model. Table~\ref{tab:app1_portmanteau} reports the test statistics at lags 6, 8, 10, and 12. None of the tests rejects the null hypothesis at the 5\% level at these four horizons.

\begin{table}[htbp]
\centering
\caption{Portmanteau diagnostic tests}
\label{tab:app1_portmanteau}
\small
\setlength{\tabcolsep}{16pt}
\begin{tabular}{lcccc}
\toprule
Lag \(h\) & 6 & 8 & 10 & 12 \\
\midrule
\(Q_T(h)\) & 11.819 & 11.952 & 12.257 & 17.110 \\
\(p\)-value & 0.066 & 0.153 & 0.268 & 0.146 \\
\bottomrule
\end{tabular}
\end{table}

Finally, Table~\ref{tab:app1_spillover_score} reports the restricted score tests developed in Section~\ref{sec:spillover} for the currency group mode. None of the three null hypotheses of zero spillovers is rejected at the 5\% level. Using the estimates in Table~\ref{tab:app1_qmle} and the definition in \eqref{eq:spillover-intensity}, the estimated ARCH and GARCH spillover intensities are 0.02932 and 0.00012, respectively. The off-diagonal entries account for 2.93\% and 0.012\% of the squared
Frobenius norms of the estimated mode-1 ARCH and GARCH coefficient
matrices, respectively. Taken together, the score tests and estimated spillover intensities suggest weak dynamic spillovers between the currency groups over the sample period.

\begin{table}[htbp]
\centering
\caption{Spillover tests for the currency group mode}
\label{tab:app1_spillover_score}
\small
\begin{tabular}{lcrr}
\toprule
Null hypothesis & df & Statistic & {\(p\)-value} \\
\midrule
\(H_{0,1}^A:\offdiag(A_1)=0\) & 2 & 0.043 & 0.979 \\
\(H_{0,1}^B:\offdiag(B_1)=0\) & 2 & 0.200 & 0.905 \\
\(H_{0,1}^{\mathrm{dyn}}:\offdiag(A_1)=\offdiag(B_1)=0\) & 4 & 0.859 & 0.930 \\
\bottomrule
\end{tabular}
\end{table}

\subsection{Application 2: Chinese equity tensor portfolio allocation}
\label{sec:application2}
We next study a constrained minimum-variance portfolio allocation problem using daily Chinese A-share returns from December 17, 2020, to April 18, 2025. We evaluate the out-of-sample performance of covariance forecasts from the TF-BEKK model. For each day $t$, the portfolio returns form a third-order tensor $\mathcal{X}_{t}\in\mathbb{R}^{5\times 3\times 3}$, with a total of 1,050 observations. In particular, mode 1 represents the industry sector and consists of five groups obtained by aggregating Wind Level 1 industries: financial--property, resources--utilities, industrials, consumer, and innovation. Mode 2 represents firm size and consists of small, medium, and large groups formed within each sector. Mode 3 represents corporate investment and consists of groups with low, medium, and high growth in total assets formed within each sector--size group, where total asset growth is calculated from the latest and preceding accounting information.

Using information available through the end of the initial estimation period
on April 8, 2024, we consider stocks available at every quarterly review date
during that period. For each stock, firm size is measured by average log
market capitalization over the preceding 60 trading days. We average this
measure across the review dates and retain the 35 largest stocks in each
sector. The selected 175-stock universe is used to construct the historical
training portfolios and remains fixed throughout the out-of-sample period.
Missing characteristics and corporate actions are handled according to prespecified rules detailed in Section~S.4.2 of the supplement. Within this fixed universe, stocks are assigned to the size and investment groups at each quarterly review using characteristics available at that date, and the assignments are held fixed until the next review. Thus, $X_{t,i,j,k}$ is the equally weighted daily return of the stocks in sector $i$, size group $j$, and investment group $k$. The resulting tensor contains 45 portfolio returns, with three or four stocks in each portfolio.

We evaluate the one-step-ahead conditional covariance forecasts using constrained global minimum-variance portfolios \citep{markowitz1952}. At each out-of-sample date $t_0$, let $\widehat\Sigma_{t_0}$ denote the covariance forecast and let $\bm R_{t_0}=\vecop(\tensor X_{t_0})$, with $d=45$. The portfolio weights satisfy the full-investment and no-short-sale constraints and are computed as
\[
\widehat{\bm w}_{t_0}
=
\arg\min_{\bm w:\,\bm 1_d\trans\bm w=1,\ \bm w\ge \bm 0}
\bm w\trans \widehat\Sigma_{t_0}\bm w.
\]
Thus, the 45 elements of $\widehat{\bm w}_{t_0}$ are weights on the cell
portfolios. Within each
cell, the corresponding weight is allocated equally across its constituent
stocks. This comparison treats the cell portfolios as frictionlessly
rebalanced return indices, with performance measured before transaction costs
and with unrestricted rebalancing during stock suspensions.

We use a rolling estimation window of 800 trading days. The initial window runs from December 17, 2020, to April 8, 2024, and forecasts are evaluated over a 250-day out-of-sample period from April 9, 2024, to April 18, 2025. All model parameters are reestimated every 50 trading days using the most recent 800 observations, while the conditional covariance forecasts are updated daily. TF-BEKK estimation and filtering use returns centered by the entrywise means in each rolling estimation window; portfolio performance is computed from the original returns. Throughout, the factor rank for the TF-BEKK model is fixed at $(2,2,2)$ across all estimation windows. Further implementation details are reported in Section~S.4.2 of the supplement.

For comparison, we also consider competitive methods including MF-GARCH \citep{yu2025matrix}, TDCC \citep{yu2025tdcc}, variance-targeted scalar BEKK \citep{pedersen2014multivariate}, GO-GARCH \citep{van2002go}, RiskMetrics, and equal weighting. All methods are evaluated over the same out-of-sample period and under the same portfolio constraints. MF-GARCH is fitted to the matrix unfolding with sectors indexing rows and combinations of size and investment indexing columns, with implementation details given in Section~S.4.2. We use nonlinear shrinkage (NLS) \citep{ledoitwolf2012} to estimate the idiosyncratic covariance matrices in TF-BEKK and MF-GARCH models.

\begin{figure}[!htp]
\centering
\includegraphics[width=0.88\linewidth]{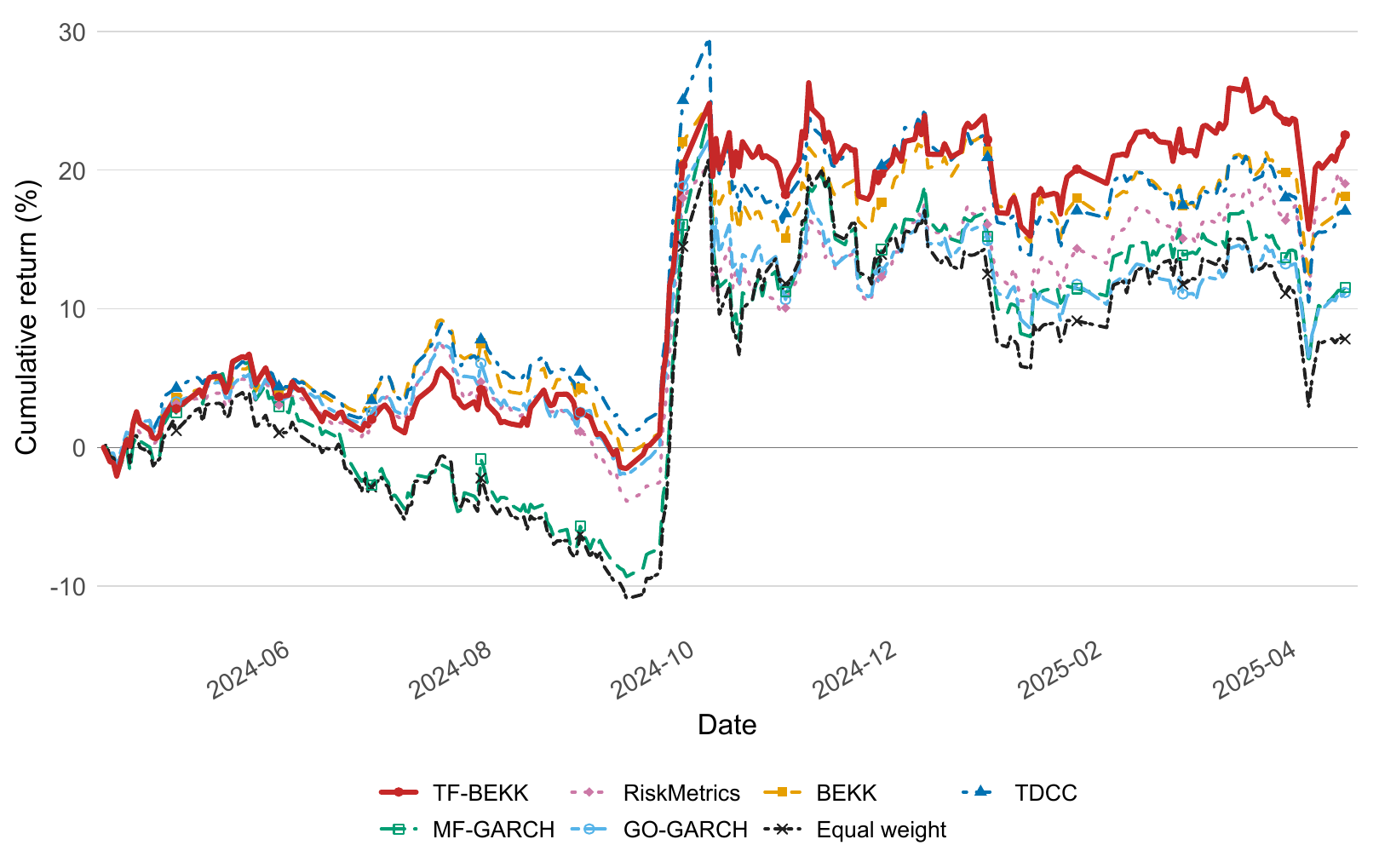}
\caption{Cumulative returns of the constrained portfolios over the 250-day out-of-sample period.}
\label{fig:app2_cumulative_returns}
\end{figure}

The annualized average return (AV), annualized volatility (SD), and information ratio $\mathrm{IR}=\mathrm{AV}/\mathrm{SD}$ for all methods are reported in Table~\ref{tab:app2_oos}.
TF-BEKK achieves the highest AV and IR over this evaluation period, at $22.21\%$ and $1.19$, respectively. MF-GARCH has the lowest SD, at $18.05\%$, compared with $18.70\%$ for TF-BEKK. Figure~\ref{fig:app2_cumulative_returns} compares the cumulative returns of the seven portfolios over the out-of-sample period. TF-BEKK ends the evaluation period with the highest cumulative return.

\begin{table}[!htbp]
\centering
\caption{Constrained portfolio performance over the 250-day out-of-sample period}
\label{tab:app2_oos}
\small
\begin{tabular*}{0.8\linewidth}{@{\extracolsep{\fill}}l
S[table-format=2.2] S[table-format=2.2] S[table-format=1.2]}
\toprule
Model & {AV} & {SD} & {IR} \\
\midrule
TF-BEKK      & 22.21 & 18.70 & 1.19 \\
MF-GARCH     & 16.23 & 18.05 & 0.90 \\
RiskMetrics  & 19.16 & 18.87 & 1.02 \\
Scalar BEKK  & 18.48 & 18.31 & 1.01 \\
TDCC         & 18.40 & 19.61 & 0.94 \\
GO-GARCH     & 12.38 & 18.50 & 0.67 \\
Equal weight & 10.19 & 22.86 & 0.45 \\
\bottomrule
\end{tabular*}
\end{table}

\section{Conclusion}
\label{sec:conclusion}

This paper develops T-BEKK, a tensor-structured BEKK model for conditional
covariance dynamics in tensor-valued time series. By imposing Kronecker
structure on the BEKK intercept and dynamic coefficient matrices, the model
reduces dimensionality while preserving a positive definite covariance
recursion and a mode-specific interpretation of the covariance intercept, ARCH
effects, and GARCH persistence.

We establish the main probabilistic, identification, and QMLE theory for
T-BEKK, and develop mode-specific spillover tests and a portmanteau diagnostic
for model checking. The TF-BEKK extension further combines tensor factor
reduction with a latent T-BEKK recursion for higher-dimensional settings. Prespecified reference matrices fix the factor orientation for parameter estimation and inference.
Simulations examine the finite-sample behavior of the estimation and inference
procedures, while the applications illustrate currency covariance modeling and
portfolio allocation.

Several directions for future research follow from this framework. First, the
T-BEKK recursion could be extended to accommodate asymmetric volatility
responses and higher-order dynamics while preserving its modewise
interpretation. Second, when some mode dimensions are large, sparsity or
low-rank restrictions on the mode-specific ARCH and GARCH matrices may provide
further dimension reduction. Finally, future research could consider time-varying factor structures
that allow the loading spaces and factor ranks to evolve, thereby accommodating
structural changes in tensor-valued systems. Extending the portmanteau result
to a growing lag order as in \citet{hong1996} is another direction for future research.
The proposed framework provides parsimonious conditional covariance modeling
and mode-specific volatility spillover inference for tensor-valued time series.

\section*{Supplementary Material}

The supplement collects the auxiliary lemmas and proofs for the main T-BEKK
and TF-BEKK results, detailed QMLE Monte Carlo results, and additional
implementation details and diagnostics for the two empirical applications.







\linespread{1}\selectfont
\bibliographystyle{apalike}
\bibliography{ref}

\end{document}